%% file: ZZ_prd.tex
\documentclass[tightenlines,twoside,secnumarabic,
               onecolumn,floatfix,nofootinbib,preprint,11pt]{revtex4}

\usepackage{graphicx}
\usepackage{dcolumn}
\usepackage{float}
\usepackage{amsmath}
\usepackage{amsfonts}
\usepackage{multirow}
\usepackage{color}
 
\begin{document}
\input macro.tex

\title{Bounding the Higgs width at the LHC using full analytic results for $gg \to e^+e^- \mu^+ \mu^-$.}

\author{John M. Campbell}
\email{johnmc@fnal.gov}
\affiliation{Fermilab, Batavia, IL 60510, USA}
\author{R. Keith Ellis}
\email{ellis@fnal.gov}
\affiliation{Fermilab, Batavia, IL 60510, USA}
\author{Ciaran Williams}
\email{ciaran@nbi.dk}
\affiliation{Niels Bohr International Academy and Discovery Center,
The Niels Bohr Institute, Blegdamsvej 17, DK-2100 Copenhagen \O, Denmark}
\preprint{FERMILAB-PUB-13-508-T}

\begin{abstract}
We revisit the hadronic production of the
four-lepton final state, $e^- e^+ \mu^- \mu^+$, through the fusion of
initial state gluons.  This process is mediated by loops of quarks and we provide
first full analytic results for helicity amplitudes 
that account for both the effects of the quark mass in the loop
and off-shell vector bosons.  The  analytic results have been implemented in the
Monte Carlo program MCFM and are both fast, and numerically
stable in the region of low $Z$ transverse momentum.  We use our
results to study the interference between Higgs-mediated and continuum
production of four-lepton final states, which is
necessary in order to obtain accurate theoretical predictions
outside the Higgs resonance region.  We have confirmed and extended a recent
analysis of Caola and Melnikov that proposes to use a measurement of the off-shell
region to constrain the total width of the Higgs boson.
Using a simple cut-and-count method, existing LHC data should bound the width at
the level of $25$--$45$ times the Standard Model expectation.
We investigate the power of using a matrix element method to construct a kinematic
discriminant to sharpen the constraint. 
In our analysis the bound on the Higgs width is improved by a factor
of about $1.6$ using a simple cut on the MEM discriminant, compared to
an invariant mass cut $m_{4l}> 300$~GeV.
\end{abstract}
\maketitle

\section{Introduction}
The discovery of a boson consistent with the Standard Model
Higgs~\cite{Aad:2012tfa,Chatrchyan:2012ufa} has set a large part of
the agenda for the LHC physics program over the next couple of
decades. The data collected in Run 1 has provided first
information about the new particle. The mass of the new
boson has been measured to be near 126
GeV~\cite{ATLAS:2013mma,Chatrchyan:2012jja} and the $0^+$ spin-parity
state is strongly favoured~\cite{Aad:2013xqa,Chatrchyan:2012jja}.
Finally, the total rates of production and decay of the boson are
broadly compatible with the predictions of the Standard Model~\cite{Aad:2013wqa,CMS:yva}.

Turning the observed cross sections into statements regarding the
coupling of the Higgs boson to Standard Model particles is a
non-trivial, but desirable goal. A typical measurement of a Higgs
process at the LHC focuses on events which lie in the Higgs
resonance region, where the cross section depends on the initial and
final state Higgs couplings, $g_i,g_f$, and on the total width as follows,
\begin{eqnarray}
\sigma_{i\rightarrow H \rightarrow f} \sim \frac{g_i^2g_f^2}{\Gamma_H}. 
\label{eq:xs_nwa}
\end{eqnarray}
Therefore in order to measure the Higgs couplings $g_{i,f}$ one must
either first measure the width, or measure the couplings under the
assumption of a known total width. Clearly, the cross section in the
narrow width approximation is invariant under the rescaling 
$g_{x} \rightarrow \xi g_{x}$ $\Gamma_{H}\rightarrow \xi^4 \Gamma_{H}$.
Information on the couplings alone can only be obtained by either
constraining the width directly, or by using ratios of cross sections
to eliminate the dependence on the total width.
Direct measurement of the Higgs width in a hadronic environment is curtailed for 
widths smaller than the detector resolution (typically
around 1 GeV). Lepton colliders offer more promising prospects,
although an $e^+e^-$ machine will only be able constrain the total
width by measuring the invisible branching fraction (in $ZH$
production). Muon colliders offer the possibility to measure the width
directly, by performing a threshold scan around the Higgs mass.

In an interesting recent paper, Caola and Melnikov~\cite{Caola:2013yja} 
proposed to constrain the total width using the number of $ZZ$ events away from the Higgs
resonance region.  This method exploits the fact that at least $15\%$
of the Higgs cross section with the Higgs boson decaying to four charged leptons
comes from the off-peak region corresponding to a four-lepton invariant mass
above 130~GeV~\cite{Kauer:2012hd}.
In the phase space region away from the Higgs 
resonance Eq.~(\ref{eq:xs_nwa}) is no longer valid, since the Higgs propagator 
is dominated by the $(s-m_H^2)$ term for large $s$ and the cross section is essentially
independent of the width. Therefore if one performs 
the same rescaling  
$g_{x} \rightarrow \xi g_{x}$, $\Gamma_{H}\rightarrow \xi^4 \Gamma_{H}$ 
the compensation which occurs in the resonance region no longer exists. 
The off-shell cross section thus depends on $\xi$ and 
therefore by measuring the total number of off-shell Higgs events one can place a
limit on the total width.
The method proposed in ref.~\cite{Caola:2013yja} using Run I data 
suggests current constraints on the total width corresponding to
$\Gamma_{H} \lesssim (20-38) \Gamma_{H}^{SM}$, with a potential limit of around
$\Gamma_{H} \lesssim (5-10) \Gamma_{H}^{SM}$ obtainable with larger LHC data sets
and sufficient control of experimental and theoretical systematic uncertainties.

\begin{figure}
\begin{center} 
\includegraphics[angle=270,width=0.8\textwidth]{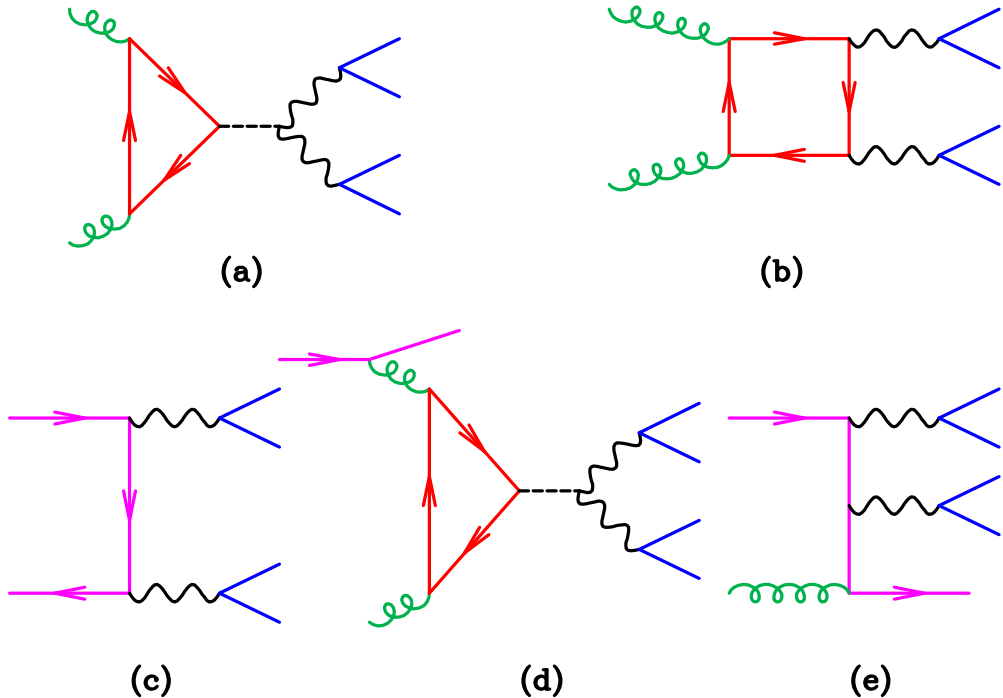} 
\caption{Representative diagrams for the partonic processes considered in this paper.}
\label{alldiags}
\end{center}
\end{figure} 
In the approach of Caola and Melnikov it is imperative to  obtain a precise prediction
for the off-peak cross section.  For large $\hat{s}$ the effective field theory 
in which the top quark is integrated out is no longer valid.  
In addition, a significant contribution in this region comes from the
effect of interference between amplitudes representing the Higgs-related diagrams and those
representing the continuum background.  Example Feynman diagrams that enter the calculation
of these two amplitudes are shown in Fig.~\ref{alldiags}(a) and (b).  In the Standard Model
the impact of this interference is significant, with the result that the effect of
including the Higgs boson diagrams is to reduce rather than increase the number of off-shell events
expected. Indeed, in the SM the total number of off-shell Higgs-mediated events is negative, as the Higgs 
unitarizes the continuum $gg\rightarrow ZZ$ cross section. A similar interference effect exists in the
$H\rightarrow \gamma\gamma$ channel~\cite{Dixon:2003yb,Martin:2012xc,Martin:2013ula}
and another recent proposal~\cite{Dixon:2013haa} exploits this to similarly
constrain the total width.  This latter method requires a precise measurement of the shift in
the mass (when compared to the results in other channels such as $ZZ$) 
caused by the interference, 
to constrain the couplings of Higgs to photons and gluons.
This can then be used to constrain the total width given the form of the total cross
section formula.  An alternative to these strategies is to combine experimental results 
across all Higgs boson production and decay channels and apply extra constraints on
individual Higgs boson couplings based on theoretical arguments~\cite{Dobrescu:2012td}. 
The method of ref.~\cite{Dobrescu:2012td}
currently provides rather stringent limits on the Higgs boson width,
$\Gamma_{H} \lesssim (3-4) \Gamma_{H}^{SM}$, albeit with the
caveat of mild theoretical assumptions. 

In this paper we shall consider the hadronic production of four 
charged-leptons in the final state.  As we have already discussed, this proceeds both by the standard 
electroweak production\footnote{The extension to the case 
of identical leptons ($4e$ or $4\mu$) is easy to implement. However the effects of this interference 
are known to be small~\cite{Melia:2011tj}.},
\beq
\label{cont}
\bothdk{p + p}{Z/\gamma^*+}{Z/\gamma^*}{\mu^- + \mu^+} {e^- + e^+}
\eeq
and by the mediation of a Higgs boson produced in the $s$-channel,
\beq
\label{Higgs}
\bothdk{p + p\to H}{Z}{Z}{\mu^- + \mu^+} {e^- + e^+\, .}
\eeq
The underlying parton processes for the hadronic reactions in Eqs.~(\ref{cont}) and (\ref{Higgs}) 
are shown in Table~\ref{Partonic_processes}, $(a)$--$(c)$, with representative Feynman diagrams 
depicted in Fig.~\ref{alldiags}.
\begin{table}
\begin{center}
\begin{tabular}{|l|l|}
\hline
$(a): g(-p_1)+g(-p_2) \to H \to e^-(p_3)+e^+(p_4)+\mu^-(p_5)+\mu^+(p_6)  $ & $O(g_s^2 e^4)$ \\
$(b): g(-p_1)+g(-p_2) \to e^-(p_3)+e^+(p_4)+\mu^-(p_5)+\mu^+(p_6)$   & $O(g_s^2 e^4)$ \\
$(c): q(-p_1)+\qb(-p_2) \to e^-(p_3)+e^+(p_4)+\mu^-(p_5)+\mu^+(p_6)$ & $O(e^4)$ \\
$(d): q(-p_1)+g(-p_2) \to H \to e^-(p_3)+e^+(p_4)+\mu^-(p_5)+\mu^+(p_6)+q(p_7)$ & $O(g_s^3 e^4)$ \\
$(e): q(-p_1)+g(-p_2) \to e^-(p_3)+e^+(p_4)+\mu^-(p_5)+\mu^+(p_6)+q(p_7)$ & $O(g_s e^4)$ \\
\hline
\end{tabular}
\end{center}
\renewcommand{\baselinestretch}{1}
\caption{Partonic processes which contribute to the four charged-lepton final state.
The second column shows the order of the strong coupling, $g_s$, and the electromagnetic coupling, $e$,
in which the partonic process first contributes. For the purposes of this counting we do not distinguish
between the weak coupling $g_W$, the electromagnetic coupling $e$, and the Yukawa coupling $g_W m_t/2/M_W$.
In the cases where the initial and final states are the same, interference needs to 
be taken into account.}
\label{Partonic_processes}
\end{table}
We shall refer to the amplitude for the Higgs production process $(a)$ 
in Table~\ref{Partonic_processes} as $\cM_H$ and to
the continuum amplitude $(b)$ as $\cM_C$.  The dominant continuum contribution is represented
by the quark-initiated continuum reaction $(c)$.

One of the aims of this paper is to compute the complete set of $1$-loop amplitudes for process (b),
$gg\rightarrow ZZ$,  using the spinor-helicity formalism, to provide
analytic formulae for helicity
amplitudes including massive quarks in the loop.
The amplitudes can then be included together with the Higgs-mediated diagrams
in order to provide a prediction for the number of off-shell Higgs events including
all interference effects.   The analytic results that are presented here will have a significant
advantage in calculation speed over more numerical methods. 
In addition, it is known that the amplitudes $\cM_C$,
when expressed in terms of scalar integrals, can develop 
numerical instabilities when the transverse momentum of the produced vector bosons
tends to zero. These are apparent singularities that cancel when
relations between the scalar integrals in the singular region are
taken into account. 
In calculations based on the Passarino-Veltman formalism~\cite{Passarino:1978jh}
such apparent singularities appear as inverse powers of the determinant, $\Delta_4$ of the 
Gram matrix, $G_{ij}=p_i \cdot p_j$. In particular, in our case we have 
\beqn
\Delta_4(p_1,p_2,p_{34})&=&\frac{1}{2} p_1 \cdot p_2 
[4 p_1 \cdot p_{34} \, p_2 \cdot p_{34}-2 p_1 \cdot p_2 \, p_{34} \cdot p_{34}] \nn\\
&=&\frac{1}{2} p_1 \cdot p_2 \, \spab{p_1}.{(p_3+p_4)}.{p_2} \, \spab{p_2}.{(p_3+p_4)}.{p_1} \nn\\
&=& (p_1 \cdot p_2)^2\; p_T^2
\eeqn
where $p_T$ is the transverse momentum of the vector boson with momentum $p_{34}$
and $p_1$ and $p_2$ are the momenta of the incoming partons.
These delicate numerical points are particularly
trying in this case because cuts on the transverse momenta of the
final state leptons, do not exclude the region where the $p_T$ of the
vector boson is equal to zero. Moreover, simply excising these regions 
can compromise the accuracy of the theoretical prediction.
For example, imposing a $p_T$ cut, $p_T > 7$ GeV, on the 
transverse momentum of the vector boson produced by collisions at
$\sqrt{s}=7$~TeV, would exclude $8\%$ of the $gg$-initiated cross section.
Since in a spinor helicity treatment the apparent singularities appear as $\spab{p_1}.{(p_3+p_4)}.{p_2}$,
which are proportional to the square root of the Gram determinant, the severity of the 
numerical problems is reduced\footnote{The definition of the spinor products
$\spa{i}.{j},\spb{i}.{j}$ and $\spab i.(j+k).l$ is standard; the definition
is given in Eqs.~(\ref{Spinor_products1},\ref{Spinor_products2}).}.  
Moreover the existence of a compact analytic answer allows us to 
rearrange the calculation to mitigate potential numerical problems at small $p_T$.

\begin{center} 
\begin{figure}
\includegraphics[scale=0.5,width=14cm]{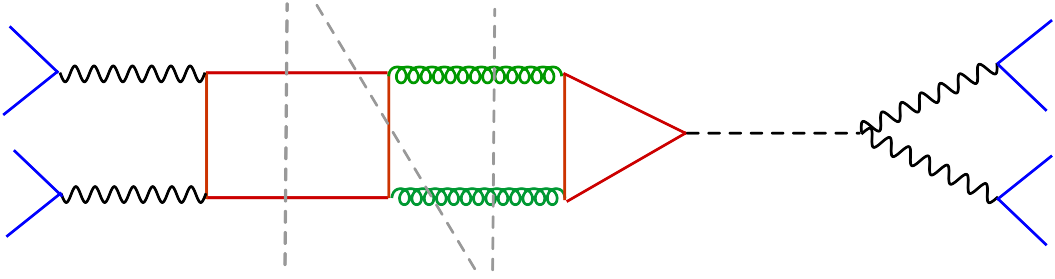} 
\caption{A representation of the
  $\mathcal{O}(g_s^4 e^8)$ interferences between Higgs
  and continuum production of four leptons considered in this paper.
  The interference contributions are obtained by cutting this diagram across
  a single dashed line.}
\label{fig:4lint}
\end{figure} 
\end{center}

Returning to Table~\ref{Partonic_processes}, we will now discuss the role of processes $(d)$ and $(e)$.
These contributions naturally arise as part of the NLO corrections to reactions $(a)$ and $(c)$
respectively.  For our purposes it is not this aspect that is most relevant.  Instead, from
the order of the couplings presented in Table~\ref{Partonic_processes}, it is clear that a
consistent treatment of Higgs-related $4$-lepton production at order $g_s^4 e^8$, i.e.
consideration of $|\cM_H + \cM_C|^2$, should also
include the interference between processes $(d)$ and $(e)$. A useful way of visualizing the
relevant interference contributions is shown in Fig.~\ref{fig:4lint} where the different contributions
are represented by various cuts of a single master topology.  In addition to the two interference
contributions highlighted here, in principle a further cut may be performed that leads to
contributions from the interference of tree-level $q\bar q \to ZZ$ and 2-loop $q\bar q \to H \to ZZ$
amplitudes. However, this vanishes for massless fermions since by helicity
conservation the $H \to q^+\overline{q}^-$ amplitude vanishes. As a
result there can only be contributions through bottom quarks, i.e. $q=b$.
We do not consider such a contribution in this paper since it is heavily suppressed
by the initial state $b$-quark parton distribution functions (PDFs)\footnote{For the same reason we also neglect potential 
cuts of the Higgs triangle loop}.
However, in order to quantify the expected number of off-shell events
expected in the four-lepton channel, we assess the impact of the $qg$
and $\overline{q}g$-initiated interferences originating from amplitudes $(d)$ and $(e)$.
These terms contain a final state parton, which may or may not be resolved as a jet, but
the contribution is finite. 

Given the importance of a measurement of the Higgs boson width, it is natural to
consider methods that could improve the limits that were suggested in Ref.~\cite{Caola:2013yja}.
One possible strategy is to use event-by-event discriminants to separate signal and background
events on a probabilistic basis.   This type of matrix element method (MEM) has already been 
successfully applied in the on-shell region~\cite{Chatrchyan:2012ufa,Chatrchyan:2012sn,Gainer:2013iya}.
In this paper we will investigate the potential of a recent MEM formulation~\cite{Campbell:2012cz}
to identify off-shell Higgs events and therefore provide more stringent constraints on the
total width of the Higgs boson.

This paper proceeds as follows.  In section~\ref{sec:hamps} we collect
the needed Higgs amplitudes for the interference studies.  In
section~\ref{sec:continuum} we discuss the calculation of the
continuum amplitudes, including an outline of the result for the
calculation of the $gg \rightarrow 4\ell$ continuum amplitude
including loops of massive fermions.  
Full details of the result for this one-loop
calculation are given in Appendices~\ref{app:LR} and \ref{app:LL}.  
In section~\ref{sec:Pheno} we
present a phenomenological study of the 4-lepton final state, including the effect of all
interferences considered here, and consider the
impact on Higgs width studies. We investigate the potential 
improvements on these constraints using the matrix element method in section~\ref{sec:MEM}. Finally in section~\ref{sec:conc}
we draw our conclusions.

\section{Glue-Glue initiated and Quark-gluon initiated Higgs amplitudes} 
\label{sec:hamps} 

In this section we describe the amplitudes appearing in Table~\ref{Partonic_processes} which
contain a Higgs boson, namely amplitudes $(a)$ and $(d)$.
Although the production of a Higgs boson through gluon fusion via a 
heavy fermion loop and its subsequent decay to four charged leptons
is well known~\cite{Georgi:1977gs}, 
for completeness and to introduce our notation, we reproduce the results in this section.
The amplitudes for continuum processes that do not involve a Higgs boson propagator will be presented 
in Section~\ref{sec:continuum}.
\subsection{Process~$(a):gg\rightarrow H\rightarrow e^-e^+\mu^-\mu^+$}
We begin by re-deriving the well-known $gg$ initiated amplitudes.
We first extract color, couplings and phases, yielding the following definition of our reduced amplitude,
\beq
{\cal A}(1_g^{h_1},2_g^{h_2},3_e^{h_3},4_{\bar{e}}^{h_4},5_\mu^{h_5},6_{\bar{\mu}}^{h_6})=
\frac{i}{16 \pi^2} \frac{\delta^{C_1,C_2}}{2} 8 e^4 g_s^2  
      \, A(1_g^{h_1},2_g^{h_2},3_e^{h_3},4_{\bar{e}}^{h_4},5_\mu^{h_5},6_{\bar{\mu}}^{h_6}). 
\eeq
Since the Higgs boson is a propagating $s$-channel scalar we can further divide this amplitude into component pieces, 
\beq
A(1_g^{h_1},2_g^{h_2},3_e^{h_3},4_{\bar{e}}^{h_4},5_\mu^{h_5},6_{\bar{\mu}}^{h_6})
 = A^{gg \to H} (1_g^{h_1},2_g^{h_2})
 \times \frac{\prop{H}(s_{12})}{s_{12}} 
 \times A^{H \to 4l}(3_e^{h_3},4_{\bar{e}}^{h_4},5_\mu^{h_5},6_{\bar{\mu}}^{h_6}) 
\eeq
where $ A^{gg \to H} (1_g^{h_1},2_g^{h_2})$ represents the Higgs
production through gluon fusion, and 
$A^{H \to 4l}(3_e^{h_3},4_{\bar{e}}^{h_4},5_\mu^{h_5},6_{\bar{\mu}}^{h_6})$
represents the decay of the Higgs into four-leptons.  The amplitudes
are sewn together using the propagator function $\prop{H}(s)$,
\beq \label{propdef}
\prop{X}(s) = \frac{s}{s-M_X^2+i M_X \Gamma_X}.
\eeq

As is well-known, for a spin zero Higgs boson there are only two non-zero helicity
amplitudes, namely those in which the two gluons have the same helicity. In
these instances the amplitude has the following form,
\begin{eqnarray}
A^{gg \to H}(1_g^+,2_g^+) &=& \frac{\spb{1}.{2}} {\spa{1}.{2}}  \Big[\frac{m^2}{2 M_W \sin \theta_W } 
\Big(2-\sud \, C_0(p_1,p_2,m,m,m) (1-\frac{4 m^2}{\sud})\Big)\Big] \nonumber \\
A^{gg \to H}(1_g^-,2_g^-) &=&  \frac{\spa{1}.{2}} {\spb{1}.{2}}  \Big[\frac{m^2}{2 M_W \sin \theta_W } 
\Big(2-\sud \, C_0(p_1,p_2,m,m,m) (1-\frac{4 m^2}{\sud})\Big)\Big] \, .
\end{eqnarray}
The function $C_0$ is the scalar triangle integral. The exact definition is given in Appendix~\ref{app:scalarintegrals}
and $m$ represents the mass of the fermion in the loop. Sizeable contributions 
result only from the cases  $m=m_t$ or $m=m_b$. 
The reduced amplitudes for decay of the Higgs boson into four leptons are defined as follows 
\beqn
\label{Htofourleptons}
A^{H \to 4l}(3_e^-,4_{\bar{e}}^+,5_\mu^-,6_{\bar{\mu}}^+)  &=& 
 \frac{M_W}{\sin \theta_W\, \cos^2\theta_W}  \frac{\prop{Z}(s_{34})}{s_{34}}\frac{\prop{Z}(s_{56})}{s_{56}} \,
\spa{3}.{5} \spb{4}.{6} \, l_e^2\nonumber \\
A^{H \to 4l}(3_e^+,4_{\bar{e}}^-,5_\mu^-,6_{\bar{\mu}}^+)  &=& 
 \frac{M_W}{\sin \theta_W\, \cos^2\theta_W}   \frac{\prop{Z}(s_{34})}{s_{34}}\frac{\prop{Z}(s_{56})}{s_{56}} \,
\spa{4}.{5} \spb{3}.{6} \, r_e l_e\nonumber \\
A^{H \to 4l}(3_e^-,4_{\bar{e}}^+,5_\mu^+,6_{\bar{\mu}}^-)  &=& 
 \frac{M_W}{\sin \theta_W\, \cos^2\theta_W} \frac{\prop{Z}(s_{34})}{s_{34}}\frac{\prop{Z}(s_{56})}{s_{56}} \,
\spa{3}.{6} \spb{4}.{5}  \, l_e r_e \nonumber \\
A^{H \to 4l}(3_e^+,4_{\bar{e}}^-,5_\mu^+,6_{\bar{\mu}}^-)  &=& 
 \frac{M_W}{\sin \theta_W\, \cos^2\theta_W }  \frac{\prop{Z}(s_{34})}{s_{34}}\frac{\prop{Z}(s_{56})}{s_{56}} \,
\spa{4}.{6} \spb{3}.{5} \, r_e^2
\eeqn
where the couplings of the $Z$ boson to the charged lepton line are,
\beqn
\label{leredef}
      l_e&=&\frac{(-1+2 \sin^2 \theta_W)}{\sin( 2\theta_W)} \\
      r_e&=&\frac{ 2 \sin^2 \theta_W}{\sin( 2\theta_W)}    
\eeqn
and $\theta_W$ is the Weinberg angle. With our conventions we recover the 
full amplitudes for Higgs decay by multiplying the expression in Eq.~(\ref{Htofourleptons}) by $-2 i e^3$.
In writing these equations we have
introduced the notation,
\beq
s_{ij} = (p_i+p_j)^2 \, ,
s_{ijk} = (p_i+p_j+p_k)^2 \,.
\eeq
We express the amplitudes in terms of spinor products defined as,
\begin{equation}   
\label{Spinor_products1}
\spa i.j=\bar{u}_-(p_i) u_+(p_j), \;\;\;
\spb i.j=\bar{u}_+(p_i) u_-(p_j), \;\;\;
\spa i.j \spb j.i = 2 p_i \cdot p_j,\;\;\;
\end{equation}     
and we further define the spinor sandwiches for massless momenta $j$ and $k$,
\beqn   
\label{Spinor_products2}
\spab i.(j+k).l &=& \spa i.j \spb j.l +\spa i.k \spb k.l \nn \\
\spba i.(j+k).l &=& \spb i.j \spa j.l +\spb i.k \spa k.l 
\eeqn

\subsection{Process~$(d):qg\rightarrow H \rightarrow e^-e^+\mu^-\mu^+ + q $ amplitudes }
For the studies of the $qg$-initiated interference we define the reduced amplitude for the 
crossed process $0\to q \qb e \eb \mu \mub g$ as follows, 
\beq
{\cal A}(1_{q}^{h_1},2_{\qb}^{h_2},3_e^{h_3},4_{\bar{e}}^{h_4},5_\mu^{h_5},6_{\bar{\mu}}^{h_6},7_g^{h_7})=
\frac{i}{16 \pi^2} \frac{1}{\sqrt{2}} \big (t^{C_7}\big)_{i_1 i_2} 8 e^4 g_s^3  
      \, A(1_{q}^{h_1},2_{\qb}^{h_2},3_e^{h_3},4_{\bar{e}}^{h_4},5_\mu^{h_5},6_{\bar{\mu}}^{h_6},7_g^{h_7}). 
\eeq
where with our conventions the reduced Higgs production amplitude is defined as,
\beq
{\cal A}(1_q^{h_1},2_{\qb}^{h_2},7_g^{h_7},H)
 = \frac{i}{16 \pi^2}  \frac{1}{\sqrt{2}} \, \big (t^{C_7}\big)_{i_1 i_2} 
   \, 4 g_s^3 e A^{q\qb g H} (1_q^{h_1},2_{\qb}^{h_2},7_g^{h_7},H)
\eeq
with ${\rm Tr}~t^{C_1}t^{C_2} = \frac{1}{2} \delta^{C_1,C_2}$. 
Since the amplitude factors onto the $s$-channel propagator in exactly
the same manner as in the previous sub-section,
\beq
A(1_q^{h_1},2_{\qb}^{h_2},3_e^{h_3},4_{\bar{e}}^{h_4},5_\mu^{h_5},6_{\bar{\mu}}^{h_6},7_g^{h_7})
 = A^{q\qb g H} (1_q^{h_1},2_{\qb}^{h_2},7_g^{h_7},H)
 \times \frac{\prop{H}(s_{127})}{s_{127}} 
 \times A^{H \to 4l}(3_e^{h_3},4_{\bar{e}}^{h_4},5_\mu^{h_5},6_{\bar{\mu}}^{h_6}) 
\eeq
the amplitudes for the decay of the Higgs $A^{H \to 4l}$ can be
re-cycled from Eq.~(\ref{Htofourleptons}).  We therefore only require
the amplitudes for production of a Higgs and $q\overline{q} g$ via a
heavy fermion loop. The two amplitudes are,
\beqn
A^{q\qb g H}(1_q^{-},2_{\qb}^{+},7_g^{+},H) &= & \frac{\spa{2}.{1} \spb{2}.{7}^2}{\sin\theta_W M_W}
       \Big[ C_0(p_{12},p_7,m,m,m) \frac{m^2}{s_{12}} \Big(\frac{1}{2}-\frac{2 m^2}{(s_{127}-s_{12})}\Big) \nonumber \\
       &+& \frac{m^2}{(s_{127}-s_{12})^2} \big (B_0(p_{12},m,m)-B_0(p_{127},m,m)\big)
       - \frac{m^2}{s_{12} (s_{127}-s_{12})}\Big] \\
A^{q\qb g H}(1_q^{-},2_{\qb}^{+},7_g^{-},H) &= & \frac{\spa{1}.{7}^2 \spb{2}.{1}}{\sin\theta_W M_W}
       \Big[ C_0(p_{12},p_7,m,m,m) \frac{m^2}{s_{12}} \Big(\frac{1}{2}-\frac{2 m^2}{(s_{127}-s_{12})}\Big) \nonumber \\
       &+& \frac{m^2}{(s_{127}-s_{12})^2} \big (B_0(p_{12},m,m)-B_0(p_{127},m,m)\big)
       - \frac{m^2}{s_{12} (s_{127}-s_{12})}\Big] 
\eeqn
The scalar integrals $B_0$ and $C_0$ are defined in Appendix~\ref{app:scalarintegrals},
and as before $m$ is the mass of the fermion circulating in the loop.

\section{Calculation of the non-Higgs boson mediated amplitudes}
\label{sec:continuum}

In this section we describe the amplitudes required for the calculation of the non-Higgs boson mediated, or continuum,
amplitudes.  These correspond to the reactions $(b)$, $(c)$ and $(e)$ in Table~\ref{Partonic_processes}.
 
\subsection{Process~$(c): q \qb \rightarrow e^-e^+\mu^-\mu^+$ }
The NLO corrections to the process,\
\begin{equation} 
q + \bar{q} \to Z Z \;,
\end{equation} 
were first calculated in refs.~\cite{Ohnemus:1990za,Mele:1990bq}, while the
inclusion of spin correlations in the decays and phenomenology for the                     
Tevatron and LHC was presented in 
refs.~\cite{Dixon:1998py,Dixon:1999di,Campbell:1999ah,Campbell:2011bn}.
This channel is the most important contribution to the four lepton production 
process. In estimating the size of this background we will use the implementation 
of this process in MCFM. This implementation includes the contributions of both 
virtual photons and $Z$-bosons in producing the final state leptons. In addition,
single resonant diagrams that contribute to the same final state are also included
through next-to-leading order.

\subsection{Process~$(b):$ Calculation of the gluon induced continuum amplitude $gg\rightarrow e^-e^+\mu^-\mu^+$ }
\label{sec:ggZZ} 
This calculation corresponds to the $gg$ initiated box diagrams, which produce pairs of
$Z$'s from a fermion loop. These calculations have a rich history. 
The first calculation of $ZZ$ production via gluon fusion (with
on-shell $Z$'s) was completed over 25 years ago~\cite{Glover:1988rg,Matsuura:1991pj}. 
These results were later
extended to include off-shell $Z$'s ~\cite{Zecher:1994kb}. More
recently, a public code {\tt{gg2VV}} was developed~\cite{Binoth:2005ua,Binoth:2008pr} 
which includes the full mass
dependence in the fermion loop, and leptonic decays of the
$Z/\gamma^*$. This code has been used to study the interference with
the Higgs signal in ref.~\cite{Kauer:2012hd}. Fully analytic helicity
amplitudes with massless fermion loops for $gg \rightarrow VV$ were presented
in ref.~\cite{Campbell:2011bn}, using the earlier results for $V+ 2j$
from ref.~\cite{Bern:1997sc}\footnote{These results were later extended to
 include the effect of the top quark mass for $gg\rightarrow
 WW$~\cite{Campbell:2011cu}. The interference with Higgs-mediated diagrams was included.}.

\begin{center} 
\begin{figure}
\includegraphics[angle=270,width=0.8\textwidth]{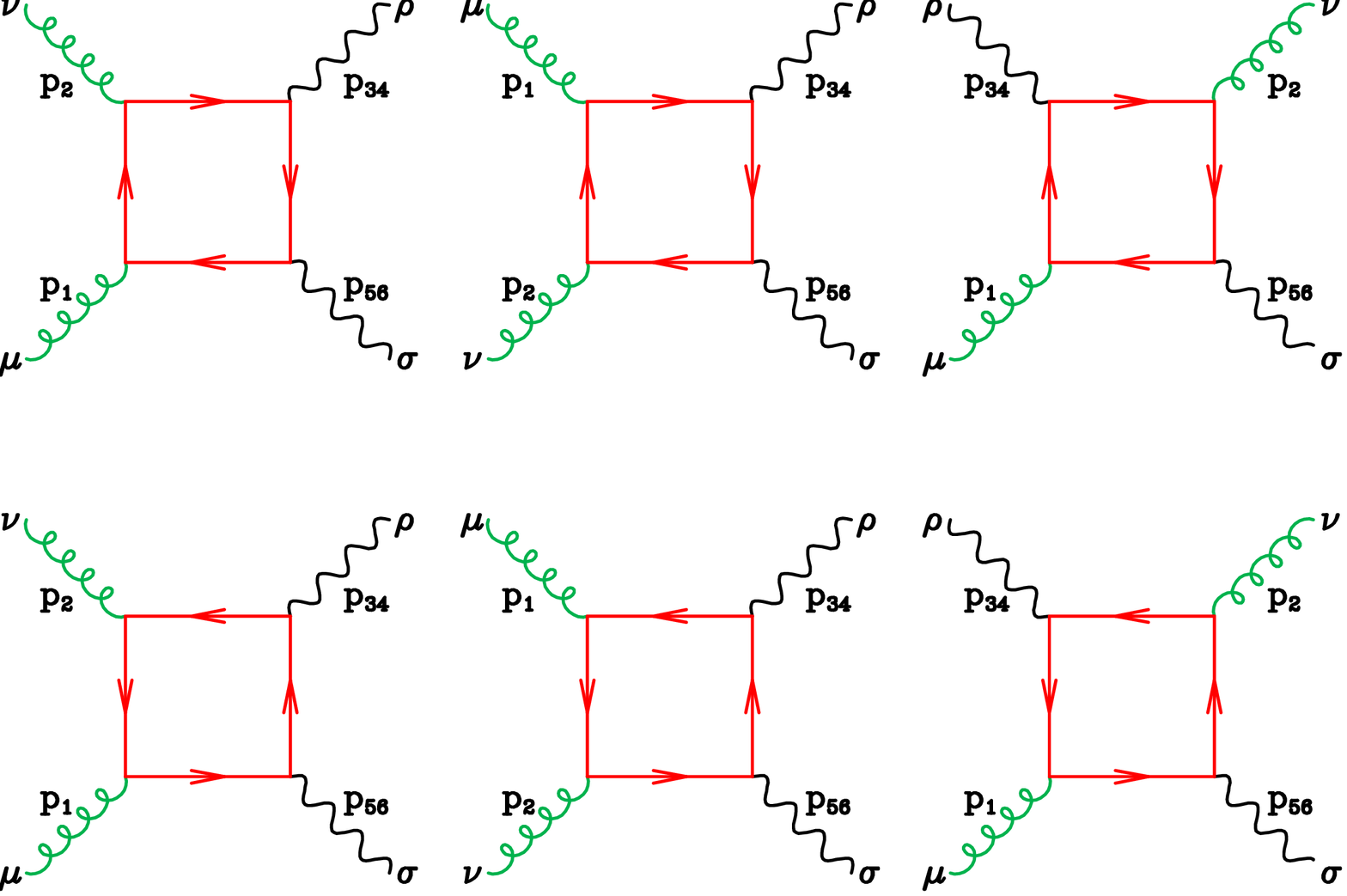} 
\caption{Diagrams for $gg \to \gamma^*\gamma^*$}
\label{ggZZ}
\end{figure} 
\end{center}

Here we will describe our analytic calculation of the helicity amplitudes for the process,
\beq
0 \to g(k_1)+g(k_2) + e^-(k_3)+e^+(k_4)+\mu^-(k_5)+\mu^+(k_6)
\eeq
with a massive fermion propagating in the loop.
The contributing diagrams are shown in Fig.~\ref{ggZZ}, where the produced electroweak
bosons that each decay to a charged lepton pair can either be a virtual photon or a $Z$-boson.
This amplitude receives contributions proportional to $V_f^2$ and $A_f^2$ (the mixed terms vanish)
where $V_f,A_f$ are the vector and axial couplings of the fermions to
the $Z$-bosons or virtual photons.

In the first instance we shall consider the leptons to be produced by an off-shell photon.
However we shall decompose the vector coupling of the photon into left- and right-handed pieces
that will be calculated separately. This is necessary for the generalization to the $Z$-boson case 
in which the left- and right-handed couplings differ. There are four sub-amplitudes 
to consider, which we denote by $LL$, $LR$, $RL$ and $RR$. The first label refers to the coupling of the
boson with momentum $p_{34}$ to the massive fermion line,
and the second to the boson with momentum $p_{56}$ to the massive fermion line. The couplings are
\beqn
&&p_{34}:\,\,\gamma^\rho   \gamma_{L/R} \nn  \\
&&p_{56}:\,\,\gamma^\sigma \gamma_{L/R}
\eeqn
where
\beq
\gamma_{R/L}=\frac{1}{2} (1 \pm \gamma_5)
\eeq
Since the mixed vector-axial contributions
vanish there are only two independent amplitudes corresponding to left-left and left-right couplings ($A_{LL}$ and $A_{LR}$)
with $A_{RR}=A_{LL}$ and $A_{RL}=A_{LR}$. Note that, if desired, the vector-vector and axial-axial contributions
can be reconstructed via,
\beq
A_{VV} = 2 \left( A_{LL}+A_{LR} \right) \;, \qquad 
A_{AA} = 2 \left( A_{LL}-A_{LR} \right) \;.
\eeq

We now describe how to construct the full amplitude containing both $Z$-bosons and virtual photons,
given the four QED amplitudes $A_{LL},A_{RL},A_{LR},A_{RR}$.
We shall make a default choice for the helicity labels of the final state leptons.
The other cases can be easily obtained by interchanging $(3 \leftrightarrow 4)$ and/or $(5 \leftrightarrow 6)$.
Our default will be to write expressions for the case
\beq \label{defaultleptonhelicity}
3^-,4^+,5^-,6^+
\eeq
In addition we will define reduced amplitudes by removing our default overall factor,
\begin{eqnarray}
{\cal A}_{jk}(1_g^{h_1},2_g^{h_2},3_e^{-},4_{\bar{e}}^{+},5_{\mu}^{-},6_{\bar{\mu}}^{+})
& =& \frac{i}{16 \pi^2}  \frac{\delta^{C_1 C_2}}{2} 8 g_s^2 e^4  \, A_{jk}
\end{eqnarray}
with $j,k=L,R$.
The full reduced amplitude for our default lepton helicity given by Eq.~(\ref{defaultleptonhelicity}) is given by,
\begin{eqnarray}
&& A(1_g^{h_1},2_g^{h_2},3_e^{-},4_{\bar{e}}^{+},5_{\mu}^{-},6_{\bar{\mu}}^{+}) = \nn \\
 &&  A_{LL}(1_g^{h_1},2_g^{h_2},3_e^{-},4_{\bar{e}}^{+},5_{\mu}^{-},6_{\bar{\mu}}^{+})
     \left( P^{L,L,-,-}(s_{34},s_{56}) + P^{R,R,-,-}(s_{34},s_{56}) \right) \nn \\
 &+& A_{LR}(1_g^{h_1},2_g^{h_2},3_e^{-},4_{\bar{e}}^{+},5_{\mu}^{-},6_{\bar{\mu}}^{+})
     \left( P^{L,R,-,-}(s_{34},s_{56}) + P^{R,L,-,-}(s_{34},s_{56}) \right)
\end{eqnarray}
The coupling factors are, for a quark of type $i$ running in the loop,
\beqn
P^{L,L,-,-}(s_{34},s_{56}) &=& (Q_i q_e +L_i l_e \prop{Z}(s_{34}))(Q_i q_e +L_i l_e \prop{Z}(s_{56})) \nn \\
P^{L,R,-,-}(s_{34},s_{56}) &=& (Q_i q_e +L_i l_e \prop{Z}(s_{34}))(Q_i q_e +R_i l_e \prop{Z}(s_{56})) \nn \\
P^{R,L,-,-}(s_{34},s_{56}) &=& (Q_i q_e +R_i l_e \prop{Z}(s_{34}))(Q_i q_e +L_i l_e \prop{Z}(s_{56})) \nn \\
P^{R,R,-,-}(s_{34},s_{56}) &=& (Q_i q_e +R_i l_e \prop{Z}(s_{34}))(Q_i q_e +R_i l_e \prop{Z}(s_{56}))
\eeqn
where
\beqn
      L_i&=&\frac{(\tau_i-2 Q_i \sin^2 \theta_W)}{\sin( 2\theta_W)} \\
      R_i&=&\frac{       -2 Q_i \sin^2 \theta_W}{\sin( 2\theta_W)}
\eeqn
and $Q_i$ and $\tau_i=\pm 1$ are the charge of the $i$th quark (in units of the positron charge) and the weak 
isospin of the $i$th quark and $l_e$ and $r_e$ are given by Eq.~(\ref{leredef}). The propagator function is defined in 
Eq.~(\ref{propdef}).

The calculational strategy for the $LR$ and $LL$ pieces will be
different. The $LR$ pieces vanish for the case of massless
quarks and consequently the tensor box integrals which occur are at most
of rank two. Because of this low rank it is easy to obtain a compact
analytic result using Passarino-Veltman reduction. Indeed the result
given in the Appendix for the $LR$ piece differs little from the result of Glover and
van der Bij~\cite{Glover:1988rg}, apart from the extension to off-shell bosons. This is
mandatory for a description of the region below the $Z$-pair threshold,
relevant for the Higgs boson. In addition we construct the helicity
amplitudes in terms of spinor products. The full result for the $LR$
helicity amplitudes is given in Appendix~\ref{app:LR}.
 
The $LL$ pieces contain tensor integrals of rank 4 and are treated with a different strategy.
For the $LL$ pieces  we use the decomposition,
\beqn
\label{Melrose_expansion}
A_{LL}(1_g^{h_1},2_g^{h_2},3_e^{-},4_{\bar{e}}^{+},5_{\mu}^{-},6_{\bar{\mu}}^{+})
& =& 
\sum_{j=2}^{3} d^{\, d=6}_j(1^{h_1},2^{h_2}) \; D_0^{d=6}(j)
+\sum_{j=1}^{3} d_j(1^{h_1},2^{h_2}) \; D_0(j) \\
&+&\sum_{j=1}^{6} c_j(1^{h_1},2^{h_2}) \; C_0(j)
+\sum_{j=1}^{6} b_j(1^{h_1},2^{h_2}) \; B_0(j) +R(1^{h_1},2^{h_2}) \nn
\eeqn
The amplitude is expanded in terms of a basis of box ($D_0$), triangle
($C_0$) and bubble ($B_0$) scalar integrals, 
with the sum running over the relevant kinematic configurations labelled
by $j$.  The precise definition of the scalar integrals is given in Appendix~\ref{app:LR}.
The basis also includes a purely rational term $R$.
The box and triangle coefficients are
determined using $D$-dimensional unitarity techniques~\cite{Britto:2004nc,Badger:2008cm, Forde:2007mi}. 
In general these coefficients are expansions in $m^2$. 
The bubble coefficients are independent of the mass and can be
constructed from the massless results of ref~\cite{Bern:1997sc}.  
There is an intimate relationship between the
$m^4$ pieces in the box coefficients, the $m^2$ pieces in the triangle
coefficients and the rational terms. We exploit these relationships
wherever possible to lighten the computational burden. The full
analytic results for the coefficients in Eq.~(\ref{Melrose_expansion})
are given in Appendix~\ref{app:LL}.

One of the features of our expansion is the introduction of the six-dimensional box
in the basis set of integrals in Eq.~(\ref{Melrose_expansion}).  We have found that the
formulation in this fashion increases the degree of numerical stability
in the low $p_T^Z$ region. The six-dimensional box can be expressed
in terms of normal four-dimensional box- and triangle scalar
integrals. This expansion introduces one power of the inverse Gram
determinant.  Note however that the apparent singularity for vanishing
Gram determinant is cancelled by relationships between the scalar
integrals in this limit. We find that grouping the terms by expressing the
four-dimensional integrals into the combination dictated by the
six-dimensional box leads to greater numerical stability.
 
\subsection{Process~$(d): q g \to e^-e^+\mu^-\mu^+ + q$}
\label{sec:qgZZ}
As before we will consider the virtual photon process first and include the additional electroweak couplings later,
\beq
{\cal A}(1_q^{h_1},2_{\qb}^{h_2},3_e^{h_3},4_{\eb}^{h_4},5_\mu^{h_5},6_{\bar{\mu}}^{h_6},7_g^{h_7}) 
   = 4 i e^4 g_s \,  \sqrt{2} \big (t^{C_7}\big)_{i_1 i_2}
A^{\rm DR/SR}(1_q^{h_1},2_{\qb}^{h_2},3_e^{h_3},4_{\eb}^{h_4},5_\mu^{h_5},6_{\bar{\mu}}^{h_6},7_g^{h_7})
\eeq
There are six Feynman diagrams for this process which can potentially contain two resonant propagators.
\beqn
&&A^{\rm DR}(1_q^-,2_{\qb}^+,3_e^-,4_{\eb}^+,5_\mu^-,6_{\bar{\mu}}^+,7_g^+) = 
         \frac{1}{\spa{1}.{7} \spa{7}.{2} s_{34} s_{56}} \nn \\
      & \times &  \Bigg\{ 
         \Bigg[\frac{\spa{5}.{1} \spab{3}.{(1+5)}.{6} \spab{1}.{(2+7)}.{4}} {s_{156}} 
          + \frac{\spa{2}.{7} \spa{5}.{1}^2 \spb{2}.{4} \spb{5}.{6} \spab{3}.{(2+4)}.{7}} {s_{234} s_{156}}  
          \Bigg]
       +  \Bigg[ 3 \leftrightarrow 5, 4 \leftrightarrow 6 \Bigg]\Bigg\}    \\
&&A^{\rm DR}(1_q^-,2_{\qb}^+,3_e^-,4_{\eb}^+,5_\mu^-,6_{\bar{\mu}}^+,7_g^-) = 
 \frac{1}{\spb{1}.{7} \spb{7}.{2} s_{34} s_{56}} \nn \\
&\times &  \Bigg\{ \Bigg[   
           \frac{\spb{2}.{6} \spab{3}.{(1+7)}.{2} \spab{5}.{(2+6)}.{4}} {s_{256}}
          + \frac{\spa{3}.{1} \spa{5}.{6} \spb{2}.{6}^2 \spb{7}.{1} \spab{7}.{(3+1)}.{4}} {s_{134} s_{256}}   
          \Bigg]  
       +  \Bigg[ 3 \leftrightarrow 5, 4 \leftrightarrow 6 \Bigg]\Bigg\}   
\eeqn
There are four Feynman diagrams for the singly resonant process which can potentially 
contain only one resonant propagator. We say `potentially' because the resonant propagators will
be added at a later stage.
\beqn   
A^{\rm SR}(1_q^-,2_{\qb}^+,3_e^-,4_{\eb}^+,5_\mu^-,6_{\bar{\mu}}^+,7_g^+) &=& 
  \frac{1}{\spa{1}.{7} \spa{7}.{2} s_{3456}} 
        \Bigg[ \frac{\spa{3}.{1} \spb{6}.{4}}{s_{56}  s_{456}}   \Big(   
                 \spab{5}.{(4+6)}.{2} \spa{2}.{1} +\spab{5}.{(4+6)}.{7} \spa{7}.{1}  \Big) \nn \\
       &+& \frac{\spa{3}.{5} \spab{1}.{(3+5)}.{6} \spab{1}.{(2+7)}.{4} }{s_{56}  s_{356}}  \Bigg]
            + \Bigg[ 3 \leftrightarrow 5, 4 \leftrightarrow 6 \Bigg]
\eeqn
\beqn
A^{\rm SR}(1_q^-,2_{\qb}^+,3_{e}^-,4_{\eb}^+,5_\mu^-,6_{\bar{\mu}}^+,7_g^-) &=&
   \frac{1}{\spb{1}.{7} \spb{7}.{2} s_{3456}} 
      \Bigg[ 
           \frac{\spa{3}.{5} \spb{2}.{4} }{s_{56}  s_{356}}   \Big(   
           \spb{2}.{1} \spab{1}.{(3+5)}.{6}
          +\spb{2}.{7} \spab{7}.{(3+5)}.{6}   \Big) \nn \\
          &+& \frac{\spb{6}.{4} \spab{3}.{(1+7)}.{2} \spab{5}.{(4+6)}.{2}}{s_{56}  s_{456}}\Bigg] 
          + \Bigg[ 3 \leftrightarrow 5, 4 \leftrightarrow 6 \Bigg]
\eeqn
The other needed helicity amplitudes can be obtained from these basic amplitudes.

\section{Phenomenology}
\label{sec:Pheno}

\begin{figure}
\begin{center} 
\includegraphics[scale=0.55,angle=270]{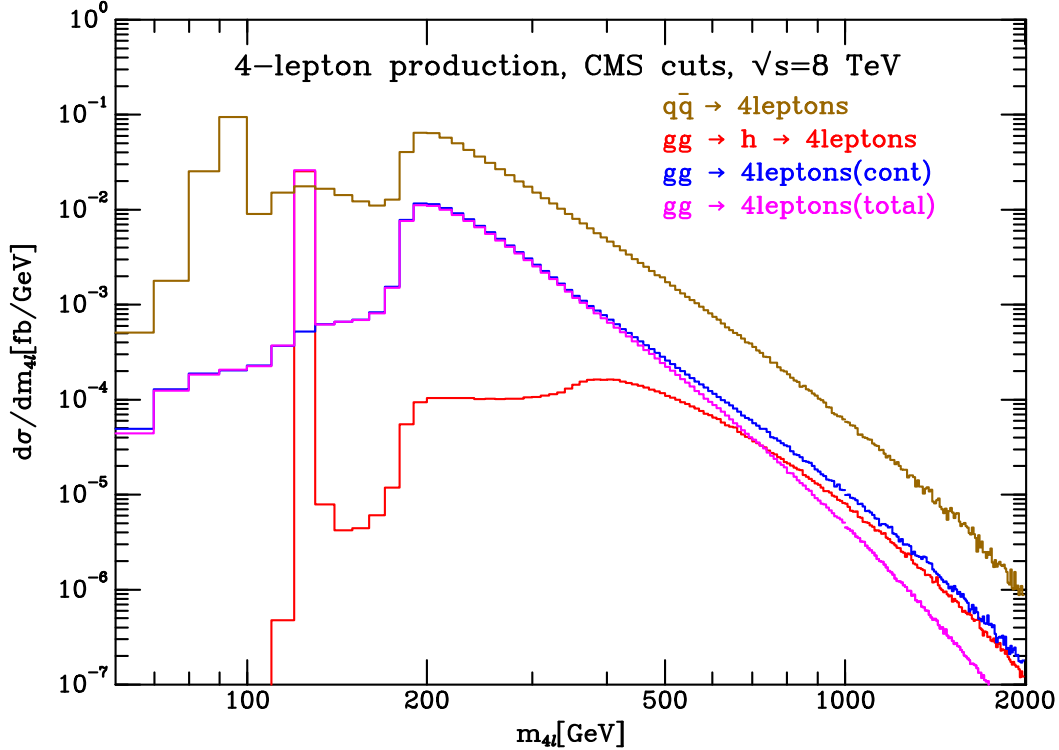} 
\caption{Overall picture at 8 TeV, (colour online). In this and the following 
figure the CMS cuts described in the text
have been imposed, but the constraint $m_{4\ell}>100~{\rm GeV}$ has been 
removed to extend the range of the plot.}
\label{Bigpicture8}
\end{center}
\end{figure}

\begin{figure}
\begin{center} 
\includegraphics[scale=0.55,angle=270]{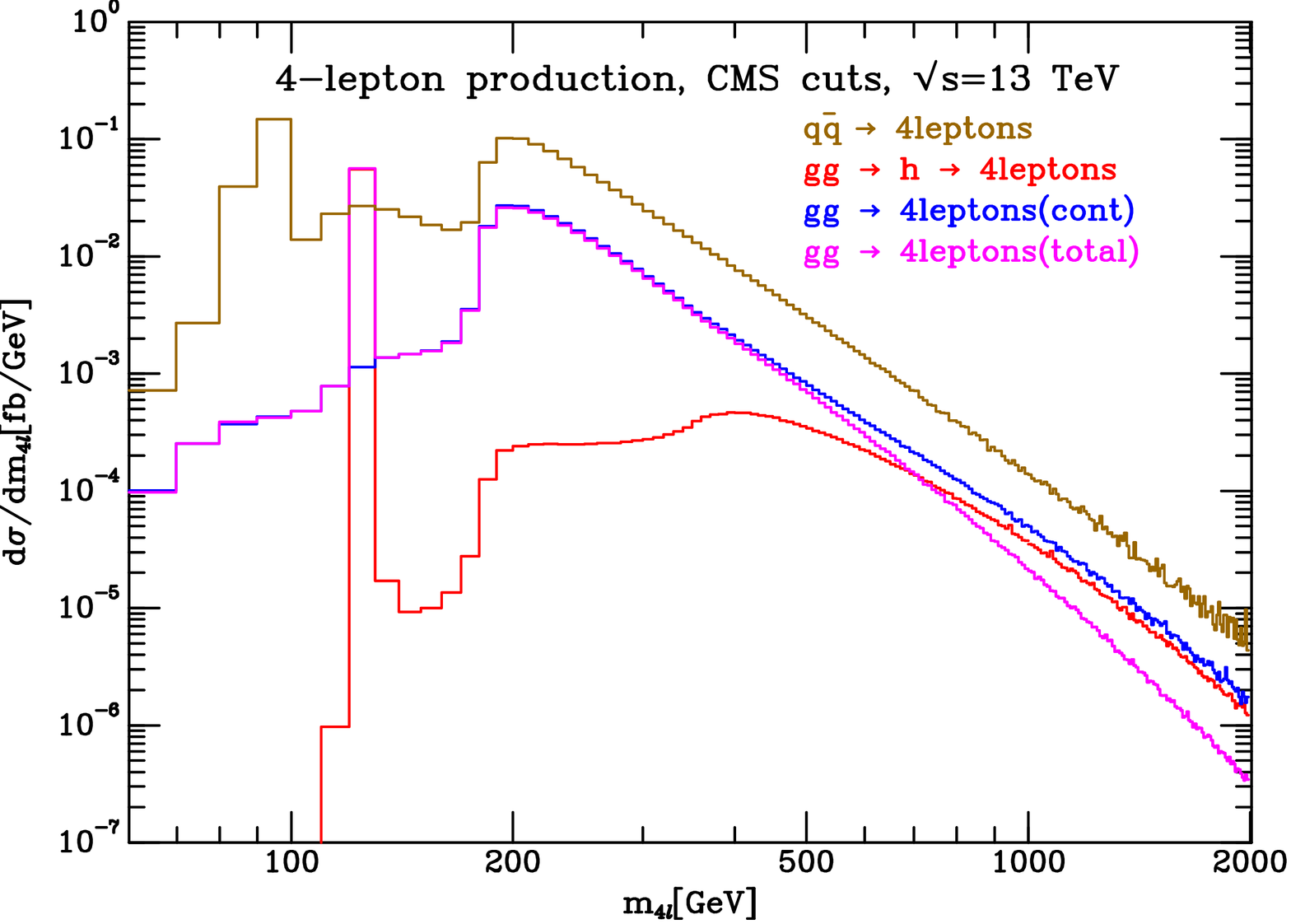} 
\caption{Overall picture at 13 TeV, (colour online).}
\label{Bigpicture13}
\end{center}
\end{figure}

The full calculation of the production of $e^- e^+  \mu^- \mu^+$ has been implemented in the 
parton level integrator MCFM. We present the relative importance of the dominant
processes at $\sqrt{s}=8$ and $13$~TeV in Figs.~\ref{Bigpicture8} and \ref{Bigpicture13}.
\begin{table}
\begin{tabular}{|c|c|c|c|}
\hline
$m_H$               & 126 GeV     & $\Gamma_H$ & 0.004307 GeV \\
$m_Z$               & 91.1876 GeV & $\Gamma_Z$ & 2.4952 GeV \\
$m_t$               & 173.2 GeV   & $m_b$      & 4.75 GeV \\
$e^2$               & 0.0949563   & $g_W^2$    & 0.4264904 \\ 
$\sin^2\theta_W$    & 0.2226459   & $G_F$      &$0.116639\times10^{-4}$ \\
\hline
\end{tabular}
\caption{Masses, widths and Electroweak parameters used to produce the results in this paper.
\label{parameters}}
\end{table}
These plots have been prepared using the parameters shown in Table~\ref{parameters} 
and applying the CMS cuts~\cite{CMS:xwa} which are detailed as follows.:
\beqn
&p_{T,\mu} > 5~{\rm GeV}\,, &  |\eta_{\mu}| < 2.4 \,, \nn \\
&p_{T,e} > 7~{\rm GeV} \,,  &  |\eta_{e}| <2.5 \,, \nn \\
&m_{ll} > 4~{\rm GeV}\,, &m_{4\ell}>100~{\rm GeV} \,.
\eeqn
In addition, the transverse momentum of the hardest (next-to-hardest) lepton should
be larger than 20 (10) GeV, 
the invariant mass of the pair
of same-flavour leptons closest to the $Z$-mass should be in
the interval $40< m_{ll}<120$~GeV and the invariant mass
of the other pair should be in the interval 
$12< m_{ll}<120$~GeV. For the purposes of these plots the QCD renormalization and factorization scales
have been set equal to $m_H/2$. 
Figs.~\ref{Bigpicture8} and \ref{Bigpicture13} contain a mixture of orders in perturbation
theory. The $q \bar{q}$ process is included at lowest order in perturbation theory $O(e^8)$,
whereas the other processes are included at $O(e^8 g_s^4)$, i.e.\ they  are next-to-next-to 
leading with respect to the $q\bar{q}$ process, but enhanced by large gluon fluxes at the LHC. 
The peaks at the Higgs boson mass and at the $Z$-boson mass (from the singly resonant diagrams)
are visible.
At high invariant mass $m_{4\ell}$ one can clearly see the destructive interference  
canceling the leading high energy behaviour of the $gg \to ZZ \to ee\mu\mu$ process~\cite{Glover:1988rg}.
Fig.~\ref{Bigpicture13} also demonstrates that the relative fraction of 
$gg$- and $q \bar{q}$-initiated processes changes at higher energy with the $gg$ process becoming 
more important at $\sqrt{s}=13$~TeV. The method of ref.~\cite{Caola:2013yja} relies on $gg$-initiated events
and is thus expected to improve with increasing energy.

\begin{center} 
\begin{figure}
\includegraphics[scale=0.55,angle=270]{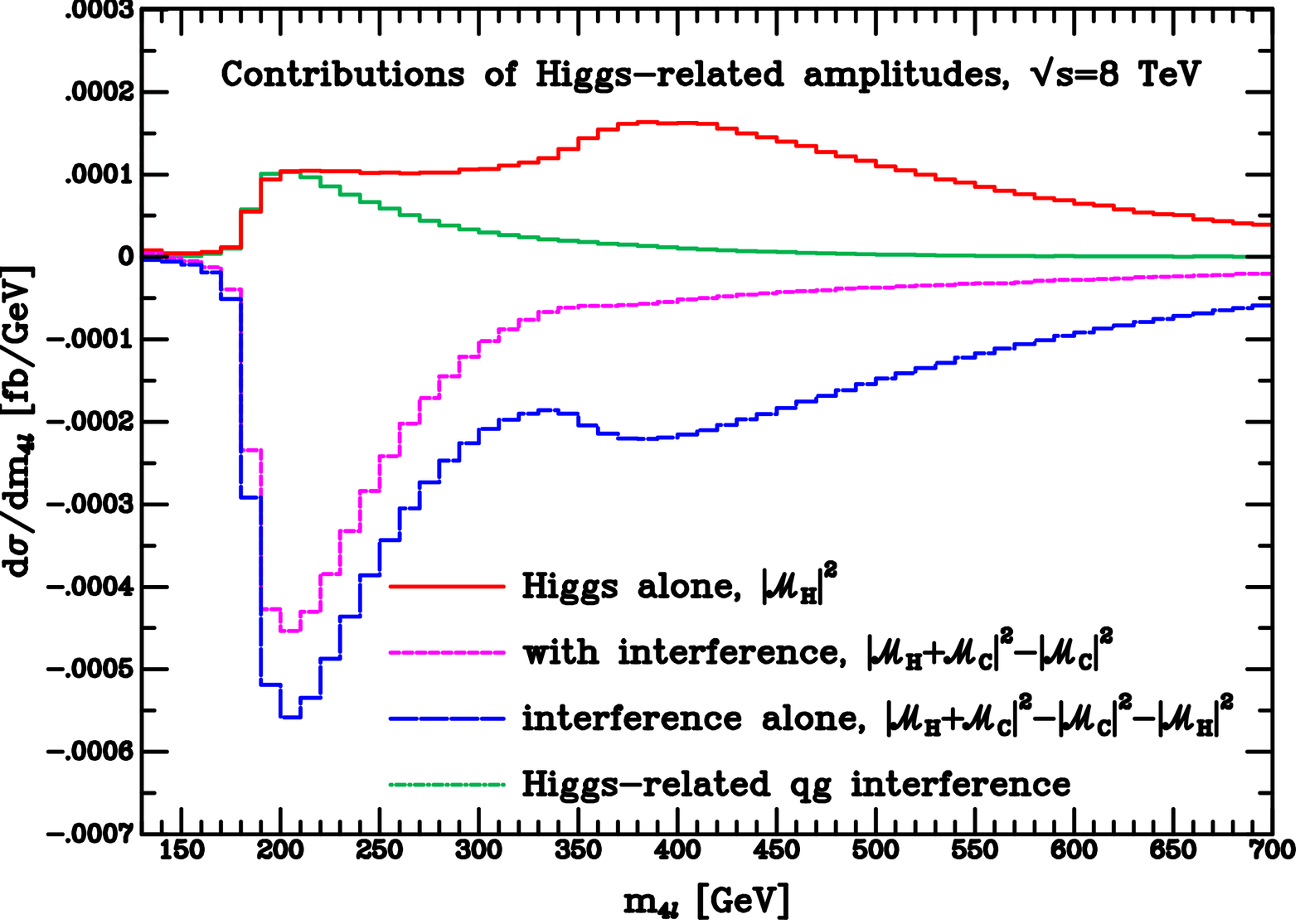} 
\caption{Higgs related contributions in the high $m_{4\ell}$ region, (colour online).}
\label{interference}
\end{figure} 
\end{center}

To discuss the structure of our results we introduce the following notation to distinguish the different
squared amplitudes that are included in the gluon-gluon initiated contributions:
\beqn
&& \sigma^H: |\cM_H|^2 \,, \qquad \sigma^C: |\cM_C|^2 \,, \qquad \sigma^{H+C}: |\cM_H+\cM_C|^2 \,, \nn \\
&& \sigma^I: |\cM_H+\cM_C|^2 - |\cM_C|^2 - |\cM_H|^2 \,,  \qquad \sigma^{H+I}: |\cM_H+\cM_C|^2 - |\cM_C|^2 \,,
\label{eq:sigmadefn}
\eeqn
where $\cM_H$ is the Higgs production amplitude and $\cM_C$ is the amplitude for the continuum background.
Thus, for instance, $\sigma^I$ reflects the pure interference contribution while $\sigma^{H+I}$ denotes
the effect of including the Higgs-mediated diagrams. 
As stressed in refs.~\cite{Caola:2013yja,Kauer:2012hd} the interference is primarily of importance in the off-peak region.
The overall size of the interference can be assessed from Fig.~\ref{interference} which shows the 
cross sections $\sigma^H$ and $\sigma^{H+I}$. It is apparent that the description of the off-peak region without
accounting for the interference is unreliable.
\begin{table}
\begin{tabular}{|c||c||c|c|c||c|c|c|}
\hline
 & $m_{4\ell}<130$~GeV &  
\multicolumn{3}{c||}{\parbox[c]{0.3\linewidth}{$m_{4\ell}>130$~GeV}} &
\multicolumn{3}{c|}{\parbox[c]{0.3\linewidth}{$m_{4\ell}>300$~GeV}} \\
Energy &   \parbox[c]{0.15\linewidth}{$\sigma^H_{peak}$}
& \parbox[c]{0.1\linewidth}{$\sigma^H_{off}$}
& \parbox[c]{0.1\linewidth}{$\sigma^{I}_{off}$}
& \parbox[c]{0.1\linewidth}{$\sigma^{qg,int}_{off}$}
& \parbox[c]{0.1\linewidth}{$\sigma^H_{off}$}
& \parbox[c]{0.1\linewidth}{$\sigma^{I}_{off}$}
& \parbox[c]{0.1\linewidth}{$\sigma^{qg,int}_{off}$} \\
\hline
\hline
7 TeV  & 0.203   & 0.044   & -0.086  & 0.0091 &0.034   & -0.050 & 0.0023 \\
\hline
8 TeV  & 0.255   & 0.061    & -0.118 & 0.011 &0.049   & -0.071 & 0.0029 \\
\hline
\hline
\end{tabular}
\caption{Fiducial cross sections for $pp \to H \to ZZ \to e^- e^+  \mu^- \mu^+ $ in fb.
All cross-sections are computed with
leading order MSTW 2008 parton distribution functions~\cite{Martin:2009iq}
and renormalization and factorization scales set equal to $m_H/2$.}
\label{CompCMTable}
\end{table}

In Table~\ref{CompCMTable} we compare our results with similar 
results presented by Caola and Melnikov.  
Our results display the same general pattern
as those reported in ref.~\cite{Caola:2013yja}, but differ in detail on the size of the 
$gg$ interference contribution, despite using what we believe to be identical input parameters.
The results of ref.~\cite{Caola:2013yja} were obtained using the code {\tt{gg2VV}}~\cite{Kauer:2012hd}.

We believe that the cause of the discrepancy is a cut of $p_T^Z>7$~GeV imposed in the double
precision version of {\tt{gg2VV}} for the continuum process, but not on the Higgs signal process.
The interference contribution is obtained by forming the combination (c.f. Eq.~(\ref{eq:sigmadefn})),
\begin{equation}
\label{real}
\sigma^I=|\cM_H+\cM_C|^2-|\cM_C|^2-|\cM_H|^2 \,.
\end{equation}
The $p_T$ cut is performed on the first two terms on the right hand side of Eq.~(\ref{real}) but not on
the third. The cut on the amplitudes that involve the continuum background in the {\tt{gg2VV}} 
code is presumably performed for reasons of numerical stability.

\begin{figure}
\begin{center} 
\includegraphics[width=8cm]{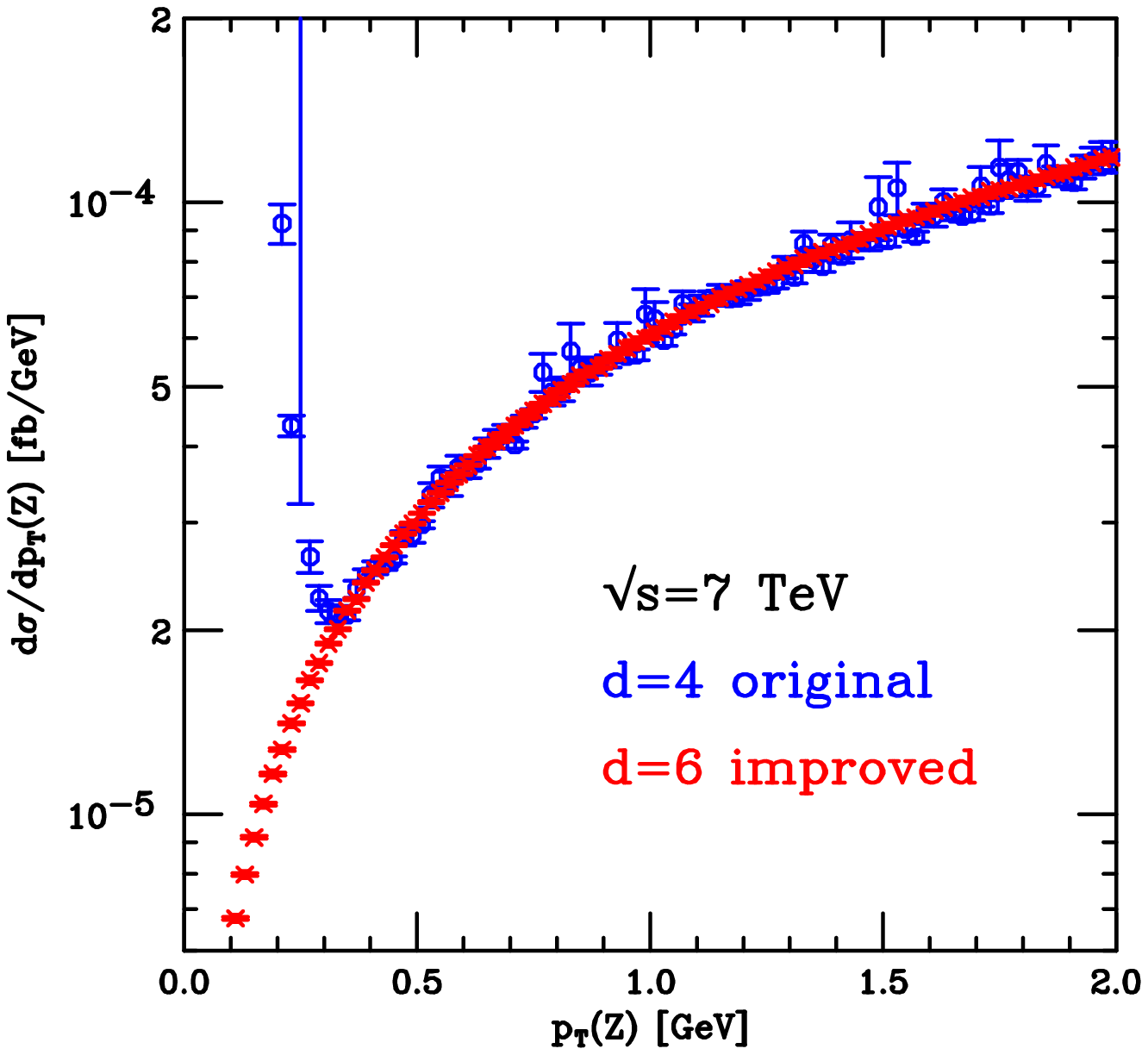} 
\includegraphics[width=8cm]{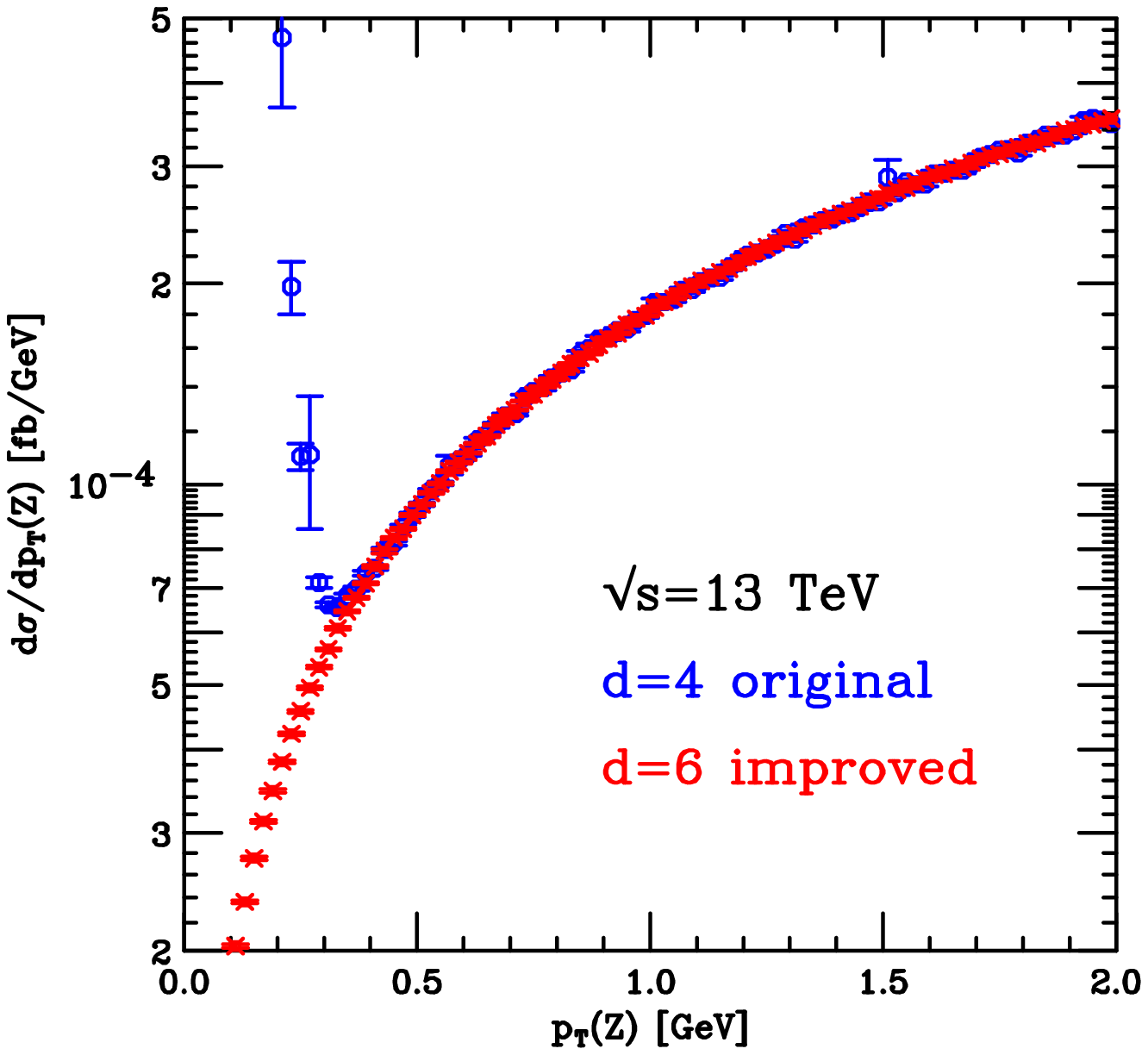} 
\caption{The $gg \to ZZ$ differential cross section, (colour online)
at $7$~TeV (left) and $13$~TeV (right), including only massive top and bottom quarks.
The calculation is performed in two different ways, as described in the text.
\label{fig:ptzcomparison}}
\end{center}
\end{figure}
\begin{figure}
\begin{center} 
\includegraphics[width=8cm]{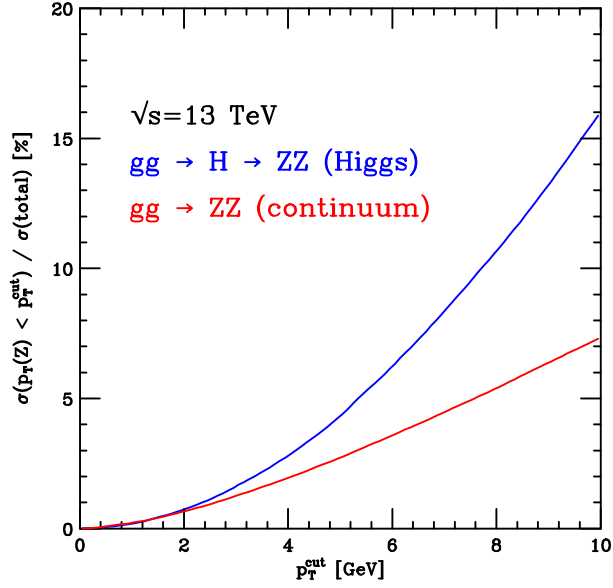} 
\caption{The percentage of the total cross section originating from the region where
the $Z$ bosons are produced with transverse momentum below a given $p_T^{\rm cut}$
at $13$~TeV. These curves are essentially the same at $\sqrt{s}=$~7 or 8 TeV.
\label{fig:xsfrac}}
\end{center}
\end{figure}

 
We shall now discuss the treatment of the region of low $p_T$ of the $Z$-boson
in our code, and illustrate the importance of low $p_T$.
In Fig.~\ref{fig:ptzcomparison} we first demonstrate the impact of the
spurious $1/p_T$ singularities that appear in the amplitudes.  The
figures show the calculation of the $gg \to ZZ$ cross section in the
region $0.1 < p_T(Z)<2$~GeV, including only the effect of the massive top
and bottom quark loops.  The calculation is performed using the CMS cuts
that were previously described.
The calculation is performed in two different
ways.  The ``original'' calculation includes only the $4$-dimensional
scalar integrals in the basis, with explicit factors of $1/p_T^4$ and
$1/p_T^3$ in the amplitudes for opposite helicity incoming gluons.  The ``improved'' calculation,
presented in Appendices \ref{app:LR},\ref{app:LL}, extends the basis to also include
$6$-dimensional box integrals, and simplifies the remaining coefficients
so that only $1/p_T^2$ factors remain. 
The original calculation becomes numerically unstable for $p_T<0.4$~GeV, whereas
the improved calculation provides a reliable prediction down to the
$p_T=0.1$~GeV threshold. 
The significance of the low-$p_T$ region is demonstrated in Fig.~\ref{fig:xsfrac}.
The figure shows the contribution to the total Higgs and continuum cross sections
from the phase space below a given $p_T$ cut.
For the $gg \to ZZ$ continuum process, the effect of enforcing a cut at $7$~GeV is a
reduction in the cross section of about $8$\%.  In contrast, a cut at the level of
$0.1$~GeV has a negligible ($<0.01$\%) effect.

In Fig.~\ref{interference} we also show the effect of Higgs-mediated diagrams 
in the $qg$($\qb g$) initiated interference and 
in Table~\ref{CompCMTable} quantify the size of this contribution to the cross section in two mass ranges.
Compared to the Higgs peak cross section this contribution is small. Further, as the figure illustrates,
this interference is significant primarily in the region around $2 m_{Z}$ and above 
$300$~GeV it can be safely neglected.  
A full assessment of the import of these
terms will have to await a complete NNLO calculation of the $pp \to e^- e^+ \mu^- \mu^+$ process since
we expect an intimate relationship between the Higgs-mediated contribution we have considered and
other box diagrams contributing to the full $q\qb gZZ$ amplitude. 
Alternatively, if we identify a jet, we could compare the data 
with a complete NLO calculation of $pp\to e^- e^+ \mu^- \mu^++{\rm jet}$, of which the interference that we present is
also a part\footnote{A NLO calculation of $pp\to ZZ+jet$ (without vector boson decays)
has been presented in ref.~\cite{Binoth:2009wk}.}.
At present we merely note that these terms do not overwhelm the contributions from the $gg$-initiated terms.
In view of the fact that our results for this interference term are small and only a partial calculation, 
we will not include them in the following discussion.

\begin{table}
\begin{tabular}{|cc|c|c|c|c|}
\hline
PDF set & Scale & \parbox[c]{0.15\linewidth}{$\sigma^H_{peak}$}
& $\sigma^H_{off}(m_{4\ell}>130$~GeV) 
& $\sigma^{I}_{off} (m_{4\ell}>130$~GeV) 
& $\left( \sigma^{H}_{off} + \sigma^{I}_{off} \right)$ / $\sigma^H_{peak}$  \\
\hline
\hline
MSTW & $m_H/2$     & 0.256 & 0.061 & -0.118 & -0.223 \\
     & $m_{4\ell}/2$  & 0.255 & 0.035 & -0.073 & -0.149 \\
\hline
CTEQ & $m_H/2$     & 0.242 & 0.052 & -0.103 & -0.252 \\
     & $m_{4\ell}/2$  & 0.243 & 0.029 & -0.065 & -0.148 \\
\hline \\
\end{tabular}
\begin{tabular}{|cc|c|c|c|c|}
\hline
PDF set & Scale & \parbox[c]{0.15\linewidth}{$\sigma^H_{peak}$}
& $\sigma^H_{off}(m_{4\ell}>300$~GeV) 
& $\sigma^{I}_{off} (m_{4\ell}>300$~GeV) 
& $\left( \sigma^{H}_{off} + \sigma^{I}_{off} \right)$ / $\sigma^H_{peak}$  \\
\hline
\hline
MSTW & $m_H/2$     &  0.256 & 0.049 & -0.071 & -0.086 \\
     & $m_{4\ell}/2$  &  0.255 & 0.026 & -0.036 & -0.039 \\
\hline
CTEQ & $m_H/2$     &  0.242 & 0.041 & -0.059 & -0.074 \\
     & $m_{4\ell}/2$  &  0.243 & 0.021 & -0.031 & -0.041 \\
\hline
\end{tabular}
\caption{Fiducial cross sections for $pp \to H \to ZZ \to e^- e^+  \mu^- \mu^+$ in fb at
$8$~TeV, with various choices of PDF sets and scale. Results are shown
for the off-peak region defined by $m_{4\ell}>130$~GeV (top) and for the
far off-peak region, $m_{4\ell}>300$~GeV (bottom).
\label{table:pdfscale}}
\end{table}
We now investigate the dependence of the on-shell, off-shell and
interference contributions on the choice of parton distribution function
and scale.  For the
sake of illustration we undertake this analysis for $\sqrt{s}=8$~TeV.
Results at other centre of mass energies are similar.
For the PDF set we consider CTEQ6L1~\cite{Pumplin:2002vw} in addition to our standard choice
of MSTW08LO~\cite{Martin:2009iq}.  We also investigate the use of a dynamic scale
that is more natural for events that lie far beyond the Higgs boson on-shell peak,
namely $m_{4\ell}/2$.  Our results are summarized in
Table~\ref{table:pdfscale}. The cross section changes considerably when switching
from the fixed to the dynamic choice of scale, since the off-peak contribution
is considerably suppressed by the running of the strong coupling.  However the
ratio of off-peak to on-peak cross sections is relatively stable under PDF variation.
Our best predictions for the effect of the interference, obtained using the running scale $m_{4\ell}/2$,
are presented in Table~\ref{BestPredictionTable}.
\begin{table}
\begin{tabular}{|cc||c||c|c||c|c|}
\hline
 &  &  &
\multicolumn{2}{c||}{\parbox[c]{0.3\linewidth}{$m_{4\ell}>130$~GeV}} &
\multicolumn{2}{c|}{\parbox[c]{0.3\linewidth}{$m_{4\ell}>300$~GeV}} \\
Energy & PDF &  \parbox[c]{0.15\linewidth}{$\sigma^H_{peak}$}
& \parbox[c]{0.15\linewidth}{$\sigma^H_{off}$}
& \parbox[c]{0.15\linewidth}{$\sigma^{I}_{off}$}
& \parbox[c]{0.15\linewidth}{$\sigma^H_{off}$}
& \parbox[c]{0.15\linewidth}{$\sigma^{I}_{off}$} \\
\hline
\hline
7 TeV  & MSTW & 0.203 & 0.025 & -0.053 & 0.017 & -0.025 \\
       & CTEQ & 0.192 & 0.021 & -0.047 & 0.015 & -0.021 \\
\hline
8 TeV  & MSTW & 0.255 & 0.034 & -0.073 & 0.025 & -0.036 \\
       & CTEQ & 0.243 & 0.031 & -0.065 & 0.022 & -0.031 \\
\hline
13 TeV & MSTW & 0.554 & 0.108 & -0.215 & 0.085 & -0.122 \\
       & CTEQ & 0.530 & 0.100 & -0.199 & 0.077 & -0.111 \\
\hline
\hline
\end{tabular}
\caption{Best prediction cross sections for $pp \to H \to ZZ \to e^- e^+  \mu^- \mu^+ $
in fb, obtained using the running scale $m_{4\ell}/2$ and two sets of parton distributions.}
\label{BestPredictionTable}
\end{table}

\begin{figure}
\begin{center} 
\includegraphics[width=8cm]{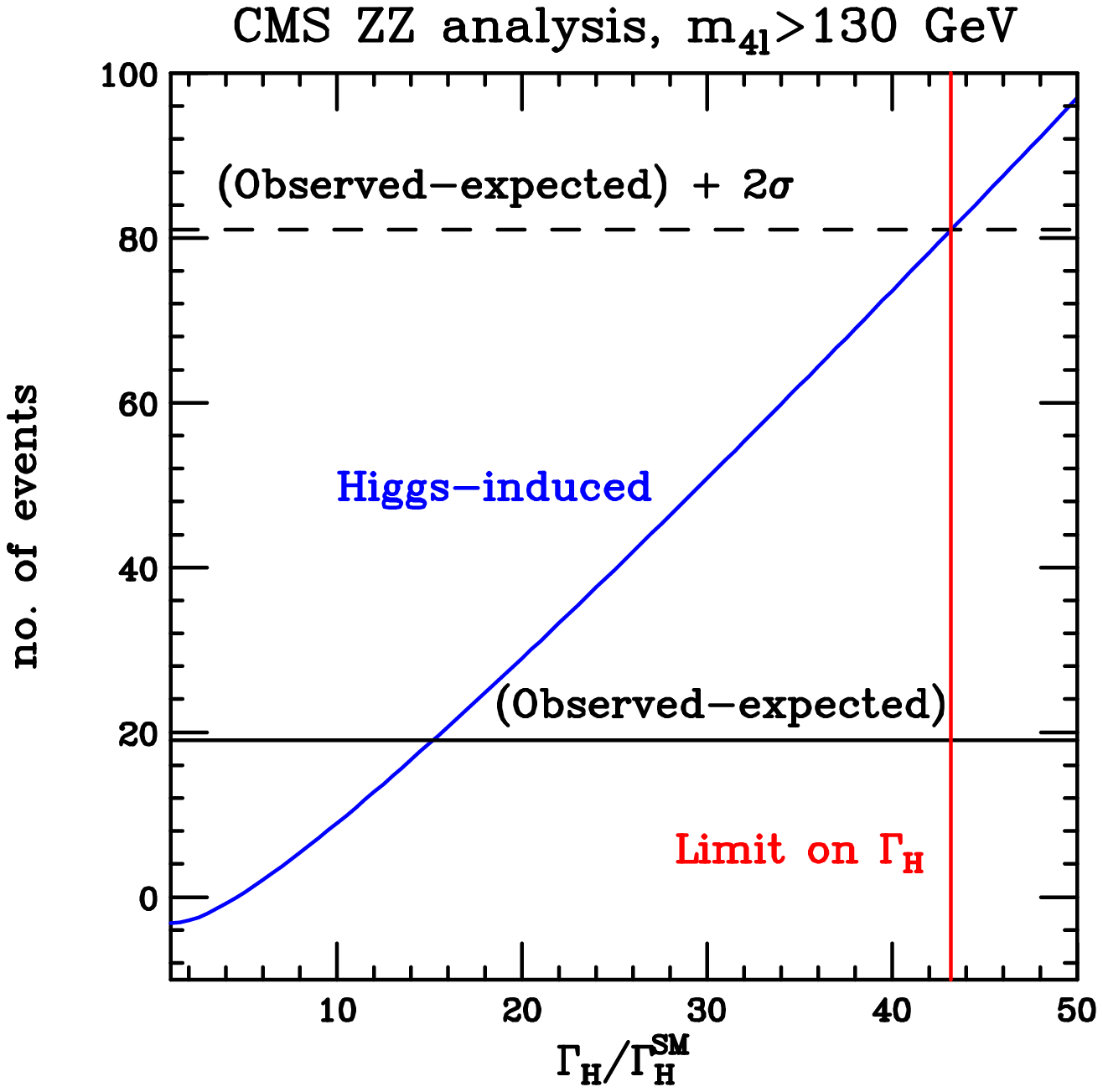} 
\includegraphics[width=8cm]{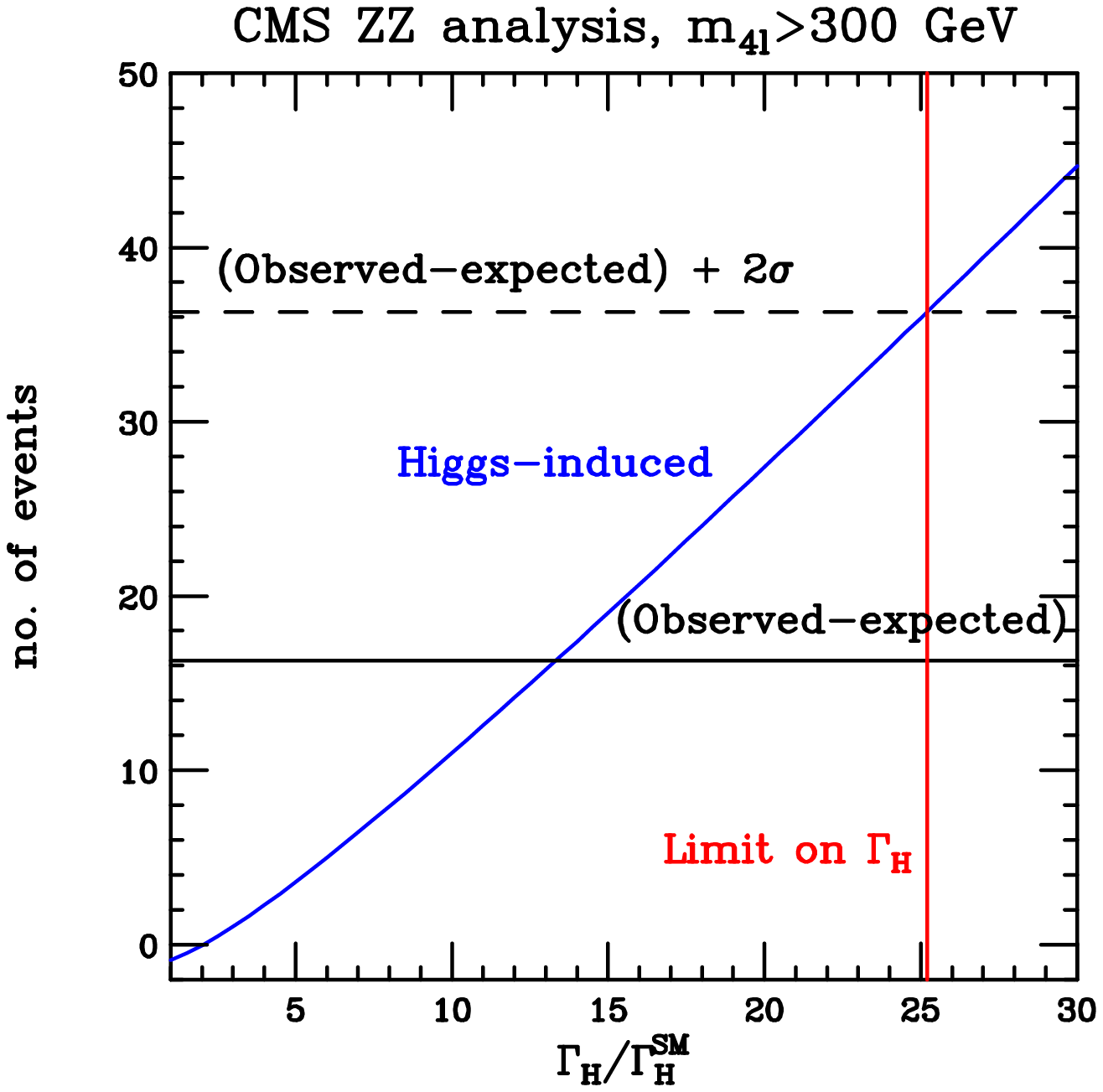} 
\caption{Limits on the width of the Higgs boson from a reanalysis of
CMS data~\protect\cite{CMS:xwa} using the method given in Ref.~\protect\cite{Caola:2013yja}.
The results of the analysis are shown for a wide off-shell region (left) and for the high-mass
region (right).
\label{fig:GammaHextraction}}
\end{center}
\end{figure}

We now turn to the issue of constraining the Higgs width by measuring the fraction of off-shell
$ZZ$ events, as proposed in Ref.~\cite{Caola:2013yja}.  The scenario we consider is one in which the
peak Higgs cross section is constrained to its Standard Model value while the width is changed.  Such
a scenario is realized by a universal rescaling of the coupling of the Higgs boson, $g_x \to \xi g^{SM}_x$
and $\Gamma_H = \xi^4 \Gamma^{SM}_H$.  Taking the results for $\sqrt{s}=8$~TeV using the MSTW PDF set from
Table~\ref{BestPredictionTable} the number of off-shell events originating from Higgs contributions is,
\begin{eqnarray}
\sigma_{off}^{H+I}(m_{4\ell}>130~\rm{GeV}) 
 &=& 0.034 \left( \frac{\Gamma_H}{\Gamma_H^{SM}} \right) - 0.073 \sqrt{\frac{\Gamma_H}{\Gamma_H^{SM}}} \\
\sigma_{off}^{H+I}(m_{4\ell}>300~\rm{GeV}) 
 &=& 0.025 \left( \frac{\Gamma_H}{\Gamma_H^{SM}} \right) - 0.036 \sqrt{\frac{\Gamma_H}{\Gamma_H^{SM}}}
\end{eqnarray}
In these equations the linear scaling with the Higgs width originates from the genuine off-shell
contribution while the interference contribution scales with the square root.  The coefficients
entering the equivalent relations at $7$ and $13$~TeV can be read directly from
Table~\ref{BestPredictionTable}.  With these results in hand it is straightforward to repeat the
analysis of Ref.~\cite{Caola:2013yja} in order to obtain the number of off-shell Higgs-related 4-lepton
events ($N_{off}^{4\ell}$) expected in the CMS analysis presented in Ref.~\cite{CMS:xwa}. The number of such events expected
in the combined $7$ and $8$ TeV data sample is obtained by summing the appropriately-weighted 
cross sections and normalizing to the peak cross section reported in  Ref.~\cite{CMS:xwa}. We find,
\begin{eqnarray}
N^{4\ell}_{off}(m_{4\ell}>130~\rm{GeV}) 
 &=& 2.78 \left( \frac{\Gamma_H}{\Gamma_H^{SM}} \right) - 5.95 \sqrt{\frac{\Gamma_H}{\Gamma_H^{SM}}} \\
N^{4\ell}_{off}(m_{4\ell}>300~\rm{GeV}) 
 &=& 2.02 \left( \frac{\Gamma_H}{\Gamma_H^{SM}} \right) - 2.91 \sqrt{\frac{\Gamma_H}{\Gamma_H^{SM}}}
\end{eqnarray}
Comparing the first of these equations to the equivalent one found in Ref.~\cite{Caola:2013yja} we see that the coefficients are
both smaller, due to the difference between our choice of dynamic scale and the approximate suppression
factor employed in Ref.~\cite{Caola:2013yja}.  The interference term differs further due to the
use of the ${\tt gg2VV}$ code in Ref.~\cite{Caola:2013yja} that employs a $p_T^{Z}$ cut, as discussed
previously.  The limit on the Higgs width is then determined by comparing the background-subtracted
number of events observed with the number of Higgs-related events expected.
This is illustrated graphically in Fig.~\ref{fig:GammaHextraction}. 
We obtain the limits,
\begin{eqnarray}
&& \Gamma_H < 43.2 \, \Gamma_H^{SM} \; \mbox{at 95\% c.l.}, \;\; (m_{4\ell}>130~{\rm GeV~analysis}) \nn \\
&& \Gamma_H < 25.2 \, \Gamma_H^{SM} \; \mbox{at 95\% c.l.}, \;\; (m_{4\ell}>300~{\rm GeV~analysis}) 
\label{eq:widthlimits}
\end{eqnarray}
These limits are slightly weaker than those reported in
Ref.~\cite{Caola:2013yja} due to the different choice of scale, as
discussed above.  Since the current limits are far from the Standard
Model value, the analysis is not affected by the small difference in
the interference term which is insignificant for large values of the
rescaling parameter.


\section{Constraining the Higgs width using the matrix element method}
\label{sec:MEM}

The results presented in the previous section highlight the difficulty
of measuring the off-shell Higgs-mediated contributions to four-lepton
production at the LHC. It is therefore natural to investigate the
possibility of using advanced techniques to extend the experimental
analyses beyond a cut and count approach. One such technique is the
use of kinematic discriminants, which assign each event a weight
associated with a given hypothesis.  The variant of this method that
we adopt is the matrix element method (MEM), in which a
fixed-order matrix element is used to assign a probabilistic weight to
individual events. In this way all of the theoretical information encoded
in the matrix element is utilized in the analysis. The MEM has been used successfully in the on-shell
region~\cite{Chatrchyan:2012ufa,Chatrchyan:2012sn,Gainer:2013iya} and
it is therefore natural to investigate the possibility of using such a
kinematic discriminant in the off-shell region.
In this section we will use the matrix element method algorithms
presented in ref.~\cite{Campbell:2012cz} to compute kinematic
discriminants in the off-shell region. Although
Ref.~\cite{Campbell:2012cz} presented an extension of the MEM to NLO
accuracy, since the $gg$ initiated matrix elements are currently only
available at LO, our analysis will focus on the LO implementation of
this algorithm. We will briefly discuss the potential impact of the
MEM@NLO at the end of this section.

The aim of the MEM is to associate a probabilistic weight to each
input event (from Monte Carlo or data), with a weight computed under
a given theoretical hypothesis.  In the case at hand we must map an
input data event to a partonic configuration in which the 4-lepton
system has no transverse momentum. In order to implement this map an
input data event, which may contain significant recoil, we perform a
transverse boost. To ensure that the
weight is unique, we integrate over all longitudinally equivalent
boosts. Each weight is thus obtained from a fixed order matrix
element, and an integration over the longitudinal degrees of freedom
associated with the production through two colliding partons.
Explicitly, at LO the weights are defined as follows,
\begin{eqnarray}
{P}_{LO}(\phi)=\frac{1}{\sigma_{LO}}\sum_{i,j} 
\int \,d x_1 dx_2 \, \delta(x_1x_2s-Q^2)
f_{i}(x_1)f_j(x_2)\hat{\sigma}_{ij}(x_1,x_2,\phi)
\end{eqnarray}
In this equation $\hat{\sigma}_{ij}$ is the LO parton cross section, evaluated at the 
phase space point $\phi$,
defined for incoming partons of flavour $i$
and $j$, which are occur in the proton with probability 
$f_{i,j}$ given by the parton distribution functions.
$Q^2$ represents the overall center of mass energy of the event that is kept invariant under the
longitudinal integration.  In this equation we have assumed that the leptons are
well-measured in order to reduce the computational load.
Lifting this assumption is straightforward and we believe
that the results presented here serve as a well-motivated and useful
starting point for future studies.

\subsection{The Kinematic Discriminant} 

For each event we compute three weights, corresponding to different hypotheses:
\begin{eqnarray}
P_{q\overline{q}} &:&\quad q\overline{q} \;  \rm{ initiated \; background}. \nonumber\\
P_{gg} &:&\quad gg \rm{\;initiated \;pieces, \;including \;Higgs \;signal, \;box \;diagrams \;and \;interference.} \nonumber\\
P_{H} &:&\quad  gg \rm{ \;initiated \;Higgs \;signal \;squared.} \nonumber 
\end{eqnarray}
The kinematic discriminant $D_S$ is then computed from these according to,
\begin{eqnarray}
D_{S} = \log{\left(\frac{P_H}{P_{gg}+P_{q\overline{q}}}\right)}
\label{eq:Disc}
\end{eqnarray}
Note that, since $P_{gg}$ contains both the effect of the Higgs diagram squared and the interference
term between the signal and background it is possible that $P_H >
P_{gg}$ so that $D_S > 0$. We have chosen $P_H$ in the numerator (compared to 
$P_{gg}$) since $P_{gg}$ will favor events which either have a large continuum  
or Higgs probability. To constrain the Higgs width we primarily seek off-shell Higgs events, and our 
discriminant is thus constructed to reflect this. 

The samples of events that we use for our study are generated as follows.
For the background $q\overline{q}$ events we use
POWHEG~\cite{Melia:2011tj} to produce NLO events matched to the
PYTHIA~\cite{Sjostrand:2006za} parton shower.  We will use the term $q\overline{q}$ background
to refer to all non $gg$-initiated backgrounds, even though this sample contains some fraction
of $gq$ initiated events that enter at NLO. Events from the Higgs
signal, $gg$ background and interference terms are generated using the
results of this paper, using the same PYTHIA interface to produce
showered events. We then perform a basic simulation of detector effects by performing Gaussian smearing of the
$p_T$ of each of the leptons, with a width of $0.5$~GeV. After this
we require exactly four leptons that pass cuts based on the CMS selection criteria presented in the previous section. For efficiency
of generation we have raised the minimum invariant mass of the off-shell lepton pair to $20$~GeV and,
for simplicity, have fixed $|\eta_{\ell}| < 2.4$ for all leptons. 

\begin{center}
\begin{figure}
\includegraphics[width=8cm]{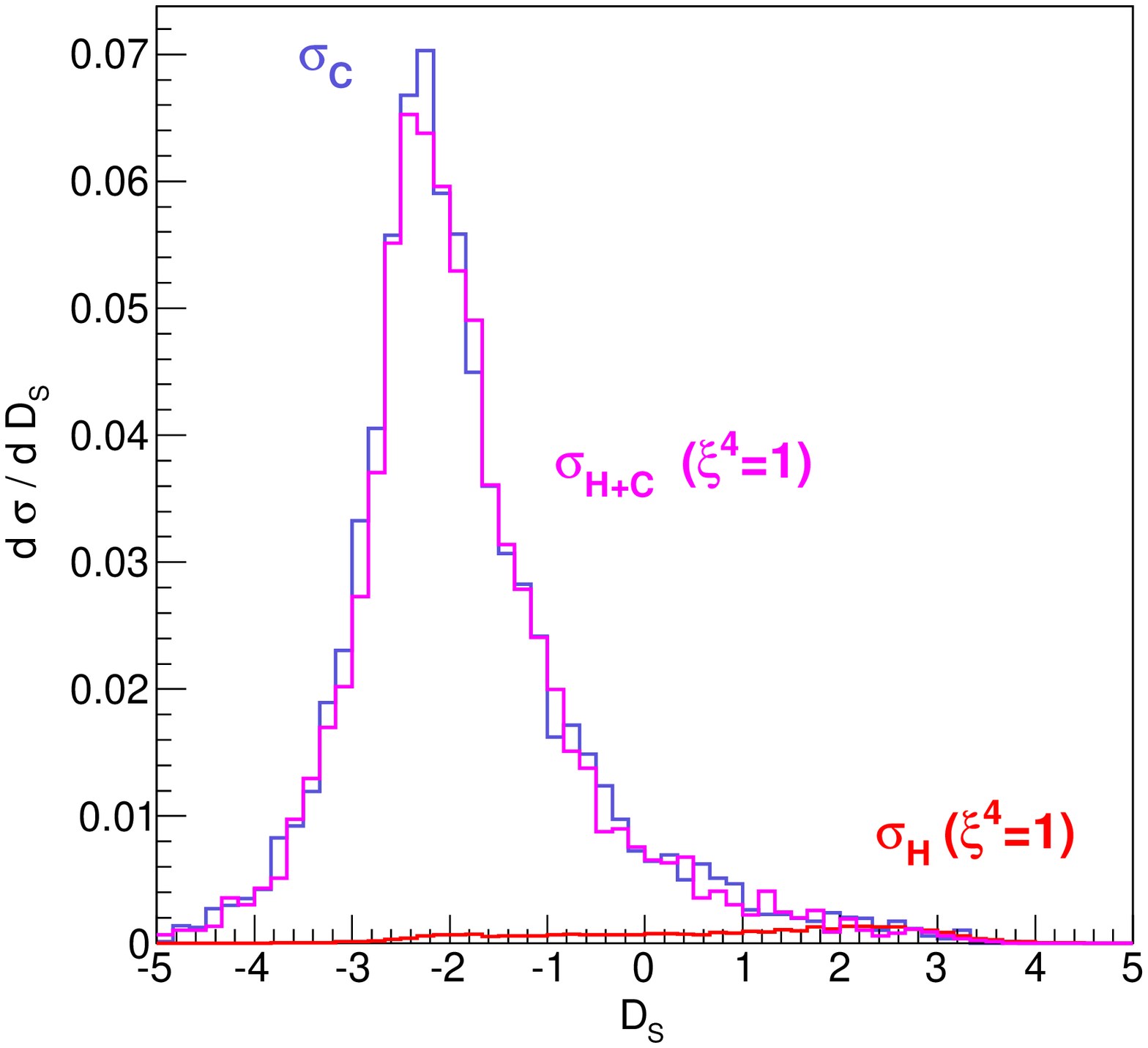}
\includegraphics[width=8cm]{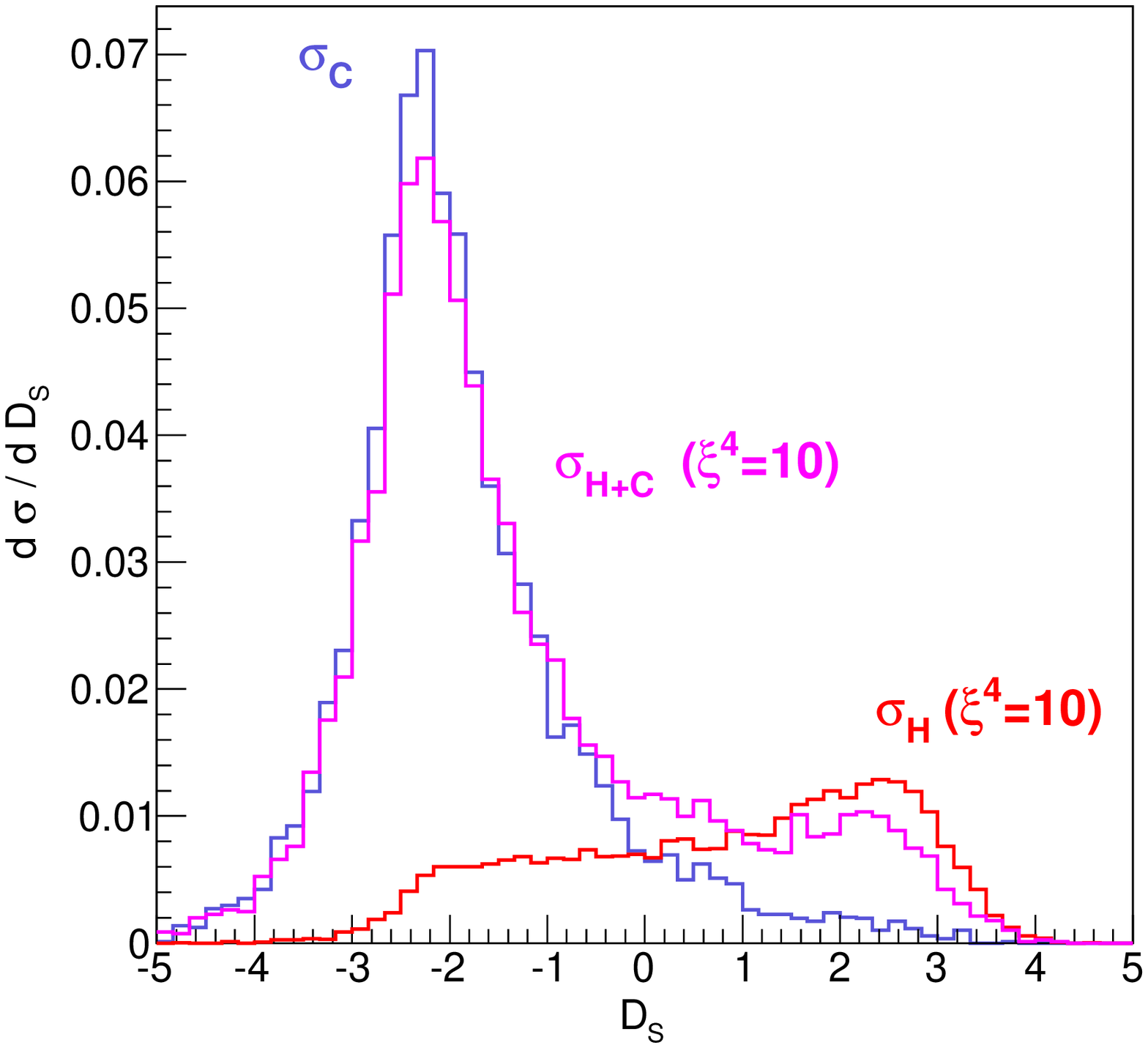}
\caption{The MEM discriminant, defined in Eq.~\ref{eq:Disc}), for 
$gg$ initiated samples corresponding to two values of the rescaling parameter $\xi$: $\xi^4=1$ (left) and
$\xi^4=10$ (right). The blue curve corresponds to the event sample containing only 
continuum production and is the same in each figure. The red curve contains only Higgs events and the magenta curve contains the full 
physical prediction, including the interference.} 
\label{fig:ggMEMval}
\end{figure}
\end{center}
We begin by validating the discriminant on our $gg$ initiated samples. Samples are generated using the prescription 
and cuts described above, for two different values of the total Higgs width:   $\Gamma_{H}\rightarrow \xi^4 \Gamma_{H}$
with $\xi^4=1,10$.  In order to understand 
the behaviour of the discriminant on the different events that may be present  we generate three samples for each $\xi$, corresponding to 
$|\mathcal{M}_{\mathcal{H}}|^2$, $|\mathcal{M}_{\mathcal{H+C}}|^2$ and $|\mathcal{M}_{\mathcal{C}}|^2$. 
Our results are summarized in Fig.~\ref{fig:ggMEMval}, which clearly indicates that the 
discriminant is working as expected.  The continuum-only sample peaks at $D_{S} \approx -2$ while the event samples containing the
Higgs boson produce a significant feature in the region $D_{S}>0$ . In addition, the number of
events present in this $D_S > 0$ region depends strongly on the rescaling factor $\xi$.  The difference between the
number of events found there between $\xi^4=1$ and $\xi^4=10$ scales roughly as $\Gamma_{H}/\Gamma_{SM}$, i.e. 
an order of magnitude. This should be compared to the overall scaling of the total $gg$ cross section, which 
for the same values of $\xi$ increases by around 24\%. 
The impact of the interference is also clear from the figure.  The destructive interference reduces the overall cross section 
and particularly suppresses the number of events in the region in which the Higgs signal is largest. These results clearly demonstrate the
importance of modeling the  interference in this measurement. Indeed, in the Standard Model the peak associated with the off-shell production 
of Higgs bosons is completely washed out by the interference, as expected from the results of the previous section.

\begin{center}
\begin{figure}
\includegraphics[width=10cm]{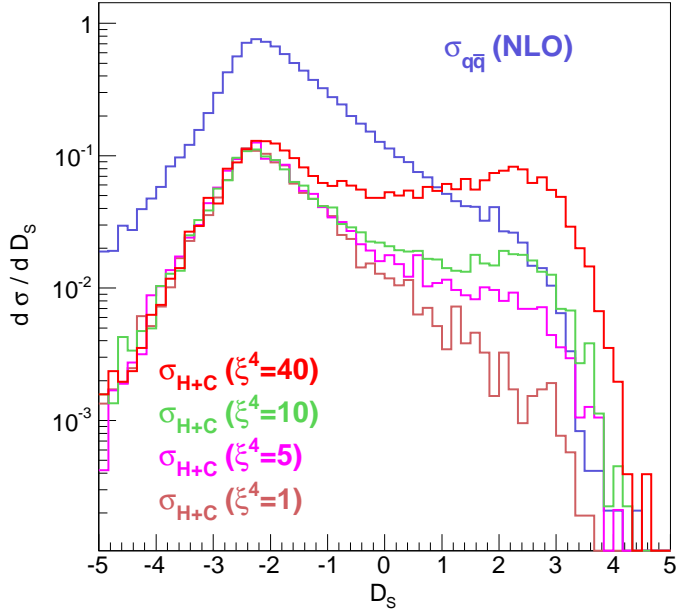}
\caption{Discriminants for the MEM (in which the discriminant is defined through Eq.~\ref{eq:Disc}) for various samples of events. 
The $q\overline{q}$ (blue) curve corresponds to the POWHEG + PYTHIA sample. The remaining curves represent four choices of the Higgs rescaling parameter $\xi$,
 corresponding to $\xi^4=1,5,10$ and 40.} 
\label{fig:DiscFull}
\end{figure}
\end{center}

Having validated our discriminant on control $gg$ samples, we now
compare our $gg$ events to the $q\overline{q}$ sample. Our results are
shown in Fig.~\ref{fig:DiscFull}.  Due to the much larger cross
section, $\sigma_{q\overline{q}}^{NLO} \approx 10 \, \sigma_{gg}$, the
$q\overline{q}$ initiated events now dominate the
discriminant. However, it is also clear from Fig.~\ref{fig:DiscFull}
that these events have the same shape as the continuum $gg$
background. As a result the region $D_S > 0$ remains sensitive to the value
of $\xi$ and for  $\xi^4 = 10$ the number of expected $gg$ events
in the tail is comparable to the number of $q\overline{q}$ events. 

\subsection{Measuring the Higgs width using the MEM} 

In order to  determine the expected limit on $\Gamma_H$ we must first form a prediction for the total number of expected events in
our data sample. In our setup the total number of expected events consists of those arising from the $q\overline{q}$, $gg$ continuum and Higgs-mediated
contributions, 
\begin{eqnarray}
 \langle N_{\rm{exp}}(\xi) \rangle =\langle N_{q\overline{q}} \rangle +\langle N^C_{gg} \rangle  + \langle N^{H+I}(\xi)\rangle
\end{eqnarray}
We wish to normalize the samples according to the number of expected $q\overline{q}$ events, i.e. we define, 
\begin{eqnarray}
 \langle N_{\rm{exp}}(\xi) \rangle =\langle N_{q\overline{q}} \rangle
 \left(1 + \frac{\sigma_{gg}^{C}}{\sigma_{q\overline{q}}} + \frac{\sigma_{gg}^{H+I}(\xi)}{\sigma_{q\overline{q}}} \right) \;.
 \label{eq:Nexpxi}
\end{eqnarray}
In Eq.~(\ref{eq:Nexpxi}) the best prediction for $\sigma_{q\overline{q}}$ 
is obtained from a NLO calculation and we generate it using POWHEG. For
$\sigma_{gg}^C$ the current state of the art is the LO calculation presented in this paper.
However the part of $\sigma_{gg}^{H+I}$ that represents Higgs diagrams squared (i.e. $\sigma_{gg}^H$)
is known to NNLO and the higher order corrections are large.
For this reason we rescale the results of this
paper for $\sigma^{H+I}_{gg}$ by a NLO $K$-factor of $1.76$.  This is derived in the effective theory, under the CMS
cuts with $m_{4\ell} > 100$ GeV.  This approach treats the higher-order corrections to the Higgs-squared diagram and the Higgs-continuum
interference equally.  However, as we have seen in the previous section, for the current LHC sensitivity the limits on the
width do not depend strongly on the effect of the interference.

In our analysis we will use a fixed $q\overline{q}$ expectation $\langle
N_{q\overline{q}} \rangle =400$. As a systematic
uncertainty on our method we will consider the variation of $\sigma_{gg}^C$ and
$\sigma_{gg}^H$ over the scale choices $\mu=\{m_{4\ell}/4,m_{4\ell}/2,m_{4\ell}\}$.
The number of Higgs-mediated events in the off-shell region, $m_{4l}>130$~GeV, can then be
parametrized by,
\begin{eqnarray} \label{Nhexp}
\langle N^H_{\rm{exp}} \rangle =
 \left\{\begin{array}{c} 2.96 \\ 2.25 \\ 1.71 \end{array}\right\}\left(\frac{\Gamma_{H}}{\Gamma_{H}^{SM}}\right)
-\left\{\begin{array}{c} 6.27 \\ 4.80 \\ 3.64 \end{array}\right\}\sqrt{\frac{\Gamma_H}{\Gamma_H^{SM}}}.
\end{eqnarray}
where each row in Eq.~(\ref{Nhexp}) corresponds to a different choice of scale.
For the statistical uncertainty we choose $N_{\rm{stat}}=1.5\sqrt{N_{\rm{exp}}}$
so that it scales correctly with the number of events and also approximately
reproduces the corresponding uncertainty in the CMS analysis~\cite{CMS:xwa}.
Without using the MEM we find, at 95\% confidence level,
\begin{equation}
\Gamma_H < \left( 41.5~{}^{-7.4}_{+10.2} \right) \, \Gamma_H^{SM} \,  \;\; (m_{4\ell}>130~{\rm GeV}) \,, \qquad
\Gamma_H < \left( 24.5~{}^{-4.9}_{+6.7} \right)\, \Gamma_H^{SM} \,  \;\; (m_{4\ell}>300~{\rm GeV}) 
\label{eq:widthlimitsmem}
\end{equation}
The systematic uncertainties in this constraint correspond to the variation of the scale about the central
value of $m_{4\ell}/2$ as described above.  Despite the small differences in the analysis compared to the last section, the
final constraints are rather similar, c.f. Eq.~(\ref{eq:widthlimits}).

\begin{center}
\begin{figure}
\includegraphics[width=10cm]{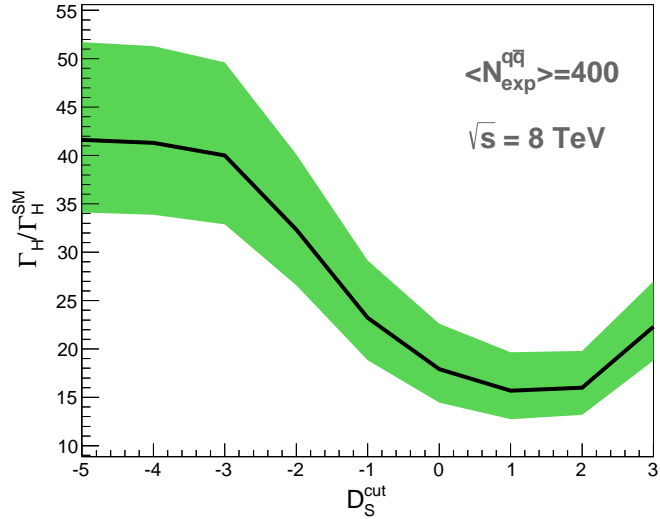}
\caption{95\% Confidence Limits on $\Gamma_H$ obtained using the Matrix Element Method. The central line corresponds 
to the limit obtained using the standard scale choices $\mu=m_{4\ell}/2$, the upper and lower limits
of the shaded band indicates the limits obtained using variations around the central scale by a factor of two. } 
\label{fig:DsGam}
\end{figure}
\end{center}

We can now compare the effect of performing a MEM analysis with a cut on the discriminant variable, $D_S > D_S^{\rm cut}$. In order
to obtain our expected number of events, given a cut on $D_S$, we use the Monte Carlo samples discussed previously (see Fig~\ref{fig:DiscFull}). For each sample we calculate the fraction of events 
that pass the cut on the discriminant. We then use the normalization prescription of Eq.~\ref{eq:Nexpxi} to combine the  
samples, weighted by the appropriate cut efficiency.  
Our results are summarized in Fig.~\ref{fig:DsGam}. It is clear
that application of a cut on the discriminant variable strengthens the constraint on the Higgs width.
Given our expected number of events, the largest values of $D_S^{cut}$ actually
result in weaker constraints on the Higgs width since there are too
few events to effectively discriminate between hypotheses. The strongest expected constraint on the Higgs width
is around $D_S^{cut} = 1$ for which we find,
\begin{equation}
\Gamma_H < \left( 15.7~{}^{-2.9}_{+3.9} \right) \, \Gamma_H^{SM} \; \mbox{at 95\% c.l.} \;.
\label{eq:widthlimitsmemfinal}
\end{equation}
This is around a factor of $2.6$ better than the cut-and-count method with $m_{4\ell} > 130$~GeV, and about 1.6 times better
than the result for $m_{4\ell} > 300$~GeV cut, c.f. Eq.~(\ref{eq:widthlimitsmem}).
Note that it may be possible to improve these limits in a full experimental analysis, for instance  by using a template fit to fully exploit 
the shape of the full $D_S$ distribution rather than simply cutting on it.

\subsection{Future Theoretical Improvements } 

The results of the previous subsection illustrate the potential of the
MEM to constrain the Higgs width.  Given its important role in
determining Higgs couplings, it is natural to consider potential
improvements which may lead to stronger constraints in the
future. Obviously the limits derived previously will improve with the
collection of larger data sets, eventually becoming dominated by
systematic errors. 

The most obvious potential improvement is the
calculation of the complete $gg$ initiated contributions (continuum
and Higgs-mediated) at NLO. This would improve both the cut and count
method, and also allow for the use of the
MEM@NLO~\cite{Campbell:2012cz}. Given the long lifetime of the LHC,
this calculation is a realistic possibility. Indeed the NLO corrections 
to the Higgs signal are already 
known~\cite{Harlander:2005rq,Anastasiou:2006hc,Anastasiou:2009kn}. 

A second improvement, that is simpler to implement, could come from binning the
events according to the number of associated jets and using the MEM@LO in
each bin separately. 
Indeed we know that for the $gg\to ZZ+\rm{jet}$ process 
the interference between Higgs and continuum diagrams in the off-peak region
is around $-160\%$ of the off-peak Higgs cross section 
and that about $9\%$ of the $gg$-initiated cross section is 
due to Higgs diagrams~\cite{Campanario:2012bh}.
This is to be contrasted with our results reported in 
Fig.~\ref{Bigpicture8}, where the interference is approximately $-200\%$ and
only about  $5\%$ of the $gg$-initiated cross section is
due to Higgs diagrams. We leave a detailed investigation of this
possibility to future work.

\section{Conclusions}
\label{sec:conc}

In this paper we have revisited the cross sections for the hadronic
production of four charged leptons, $e^- e^+ \mu^- \mu^+$, focussing
in particular on the gluon-gluon initiated process that involve closed fermion loops.  
We include the full amplitude, i.e. both the diagrams containing an $s$-channel 
Higgs boson and those proceeding through a closed loop of (massive and massless) 
fermions radiating vector bosons ($Z/\gamma^*$). Our result for the full amplitude includes the
interference between the two types of processes.  We have obtained
analytic formulae for the $gg$-initiated helicity amplitudes, retaining
the mass of the fermion that circulates in the closed loop.  Even
though numerical results using these amplitudes have been presented
before, we believe this is the first paper to publish analytic results
for helicity amplitudes including off-shell vector bosons in the final
state. The inclusion of off-shell vector bosons is clearly necessary to describe the
region where the mass of the four leptons is below twice the $Z$-boson mass,
relevant for Higgs boson studies. Our analytic approach has advantages over
a more numerical approach, both in terms of calculational speed and in terms
of numerical stability. Numerical stability can be an issue 
in the region where the vector boson transverse momentum $p_T$ is small. 
We have demonstrated that our code is stable down to $p_T=0.1$~GeV where we perform
a cut that removes a negligible fraction of the cross section (0.01\%).

The experimental study of the Higgs boson in the four lepton channel
has focussed on the resonant region where the mass of the four leptons
is close to the mass of the Higgs boson. Somewhat surprisingly, the
narrow width approximation for the Higgs boson fails because of the
proximity of the $Z$-pair threshold and the production of longitudinal
$Z$-bosons. Indeed $15\%$ of the cross section deriving from diagrams
with a Higgs boson in the $s$-channel lies hundreds of Higgs widths
above its mass, $m_{4l}>130$~GeV. It is essential to include
interference in the $gg$-channel to accurately describe this
region. The interference in the $qg$-channel is found to be smaller.
Its contribution can be further reduced by binning the data in the
number of associated jets, or by considering only the $m_{4l}>300$~GeV
region.  A definitive analysis of its importance will require a
complete higher order calculation.

Following a suggestion of Caola and
Melnikov we have used the off-resonant production through the Higgs
channel to bound the total width of the Higgs boson. We confirm and
extend the results of Caola and Melnikov, giving more precise results
for the effect of the interference and investigating alternative choices
for the renormalization and factorization scale. We find that the
choice of scale can substantially affect the ratio of off-shell to
on-shell Higgs production, although the effect of different parton
distributions on this quantity is less important. More precise
predictions for this ratio will require a campaign to include strong
and electroweak higher order effects into the cross section for the
four lepton final state.

A cut and count style analysis of current data gives limits $\Gamma_H< 43.2  \,(25.2)  \, \Gamma_H^{SM}$ 
using off-resonance events with $m_{4l}> 130 \, (300)$~GeV. 
We investigated the use of a matrix element method to construct
an event-by-event kinematic discriminant as a means of
improving the constraint on the Higgs width.
Using Monte Carlo pseudo-data we
found that such a MEM analysis could suppress 
the $q\overline{q}$ and $gg$ continuum backgrounds whilst still remaining
sensitive to the width of the Higgs boson.
In our analysis the bound on the Higgs width was improved by a factor
of about $1.6$ using a simple cut on the MEM discriminant, compared to
an invariant mass cut $m_{4l}> 300$~GeV.
Our results motivate a more complete experimental analysis 
including real data and a full detector simulation.

\noindent
\section*{Acknowledgements}
RKE would like to thank the Discovery Centre at the Niels Bohr Institute for hospitality during 
the preparation of this paper. We would like to acknowledge useful discussions with
Simon Badger, Fabrizio Caola, Andrei Gritsan and Kirill Melnikov.
The research of RKE and JMC is supported by the US DOE under contract DE-AC02-07CH11359.

\appendix

\section{Definition of scalar integrals}
\label{app:scalarintegrals}

The scalar integrals themselves are defined as follows,
\begin{eqnarray}
&& B_0(p_1;m_1,m_2)  =
 \frac{\mu^{4-d}}{i \pi^{\frac{d}{2}}\cg}\int d^d l \;
 \frac{1} {d(l,m_1) \, d(l+p_1,m_2)}
\nn \\
&& C_0(p_1,p_2;m_1,m_2,m_3)  = \frac{1}{i \pi^2}
\nn \\
&& \times \int d^4 l \;
 \frac{1} {d(l,m_1) \, d(l+p_1,m_2) \, d(l+p_1+p_2,m_3)} \\
&& D_0(p_1,p_2,p_3;m_1,m_2,m_3,m_4)
= \frac{1}{i \pi^2} \nn \\
&&
\times \int d^4 l \;
 \frac{1} {d(l,m_1) \, d(l+p_1,m_2) \, d(l+p_1+p_2,m_3) \, d(l+p_1+p_2+p_3,m_4)} \\
&&D_0^{d=6}(p_1,p_2,p_3;m_1,m_2,m_3,m_4)
= \frac{-1}{i \pi^3} \nn \\
&& \times \int d^6 l \;
 \frac{1} {d(l,m_1) \, d(l+p_1,m_2) \, d(l+p_1+p_2,m_3) \, d(l+p_1+p_2+p_3,m_4)}
\end{eqnarray}
where the denominator function is 
\beq
d(l,m)=(l^2-m^2+i\varepsilon) \, .
\eeq
For the purposes of this paper we take the masses in the
propagators to be real.  Near four dimensions we use $d=4-2 \epsilon$ (and for
clarity the small imaginary part which fixes the analytic
continuations is specified by $+i\,\varepsilon$).  
$\mu$ is a scale introduced so that the integrals
preserve their natural dimensions, despite excursions away from $d=4$.
We have removed the overall constant which occurs in $d$-dimensional integrals 
\beq
\cg\equiv\frac{\Gamma^2(1-\epsilon)\Gamma(1+\epsilon)}{\Gamma(1-2\epsilon)} = 
\frac{1}{\Gamma(1-\epsilon)} +{\cal O}(\epsilon^3) =
1-\epsilon \gamma+\epsilon^2\Big[\frac{\gamma^2}{2}-\frac{\pi^2}{12}\Big]
+{\cal O}(\epsilon^3)\,.
\eeq
The final numerical evaluation of the amplitudes 
uses the {\tt ff}~\cite{vanOldenborgh:1989wn,vanOldenborgh:1990yc}
and {\tt QCDLoop}~\cite{Ellis:2007qk} libraries
provide values for these scalar integrals.

The expression for the six-dimensional box 
with two adjacent external massless lines is, ($p_1^2=p_2^2=0$),
\beqn
\label{sixdim}
&&D_0^{d=6}(p_1,p_2,p_{34};m,m,m,m)=\frac{\sdtq}{ 2 Y} \Big[ 
\Big( \sud-\stq-\scs+2 \frac{\stq \scs}{\sdtq} \Big) C_0(p_{12},p_{34};m,m,m) \nn \\
&-&\Big(\sud \sdtq+\frac{4 m^2 Y}{\sdtq}\Big) D_0(p_1,p_2,p_{34};m,m,m,m) +\sud C_0(p_1,p_2;m,m,m) \nn \\
   &+& \sdtqm  C_0(p_2,p_{34};m,m,m) +\sucsm C_0(p_1,p_{56};m,m,m) \Big]  \, ,
\eeqn
with $Y=\sutq\sdtq-\stq\scs$.
The six-dimensional box is both infra-red and ultraviolet finite, even in the limit $m \to 0$.
We also note that, despite the overall factor of $1/Y$, it approaches a finite limit as $Y$ (or $p_T$)
goes to zero.

\section{Analytic results for the $LR$ amplitude}

\label{app:LR}
There are two independent helicity configurations,
$(1^+,2^+)$ and $(1^-,2^+)$.  To obtain the remaining helicities we define
the operation,
\beq
\label{eq:flipdef}
{\rm flip}: \qquad (3 \leftrightarrow 4), \; (5 \leftrightarrow 6) ,\; \spa{}.{} \leftrightarrow \spb{}.{} \;.
\eeq
The remaining two helicities are then obtained by,
\beqn
\label{eq:ALRflip}
A_{LR}(1_g^-,2_g^-,3_{e}^-,4_{\eb}^+,5_{\mu}^-,6_{\mub}^+)&=& {\rm flip} \left\{ A_{LR}(1_g^+,2_g^+,3_{e}^-,4_{\eb}^+,5_{\mu}^-,6_{\mub}^+) \right\} \\
A_{LR}(1_g^+,2_g^-,3_{e}^-,4_{\eb}^+,5_{\mu}^-,6_{\mub}^+)&=& {\rm flip} \left\{ A_{LR}(1_g^-,2_g^+,3_{e}^-,4_{\eb}^+,5_{\mu}^-,6_{\mub}^+) \right\} \, .
\eeqn
The $LR$ amplitude is simple because it vanishes in the $m \to 0 $ limit. 
Thus the tensor rank of the integrals that appear is at most two.
We will first consider the $LR$ amplitude for the gluonic production of two virtual photons, 
\beq
g(-p_1)+g(-p_2) \to \gamma^*(p_{34})+\gamma^*(p_{56}) \, .
\eeq
The virtual photons will subsequently decay to charged lepton pairs with momenta $p_3,p_4$ and $p_5,p_6$,
so that $p_{34}=p_3+p_4,p_{56}=p_5+p_6$. These decays will be added later in this section.

For definiteness we consider the $LR$ amplitude for the case where the couplings
of the virtual photons with momenta $p_{34}$ and $p_{56}$ are 
\beqn
p_{34}&:&\,\,-i e \gamma^\rho   \frac{1}{2} (1-\gamma_5)  \nn \\
p_{56}&:&\,\,-i e \gamma^\sigma \frac{1}{2} (1+\gamma_5) \, .
\eeqn
We remove a series of overall factors to define a reduced amplitude  $P_{LR}^{\mu \nu \rho \sigma}$
for this process,
\beq
{\cal P}^{\mu \nu \rho \sigma} = 
 \frac{i g_s^2 e^2 }{16 \pi^2} \,
 \frac{\delta^{C_1 C_2}}{2} 4 \, P_{LR}^{\mu \nu \rho \sigma} \, .
\eeq
The indices $\mu$ and $\nu$ refer to the two gluons with momenta $p_1$ and $p_2$ respectively, (see Fig.~\ref{ggZZ}).
$C_1$ and $C_2$ similarly denote the color labels of the gluons. 
Including the factor of -1 for a fermion loop we find that the form of the 
reduced amplitude, consistent with QCD gauge invariance, is~\cite{Glover:1988rg},
\beqn
P_{LR}^{\mu \nu \rho \sigma} &=&
A_1 \, g^{\rho \sigma} \Big(g^{\mu\nu}-\frac{p_1^\nu p_2^\mu}{p_1 \cdot p_2}\Big)  \nn \\
&+&A_2  \, g^{\rho \sigma} \Big( g^{\mu \nu}+\frac{2}{p_T^2}p_{34}^\mu p_{34}^\nu + \frac{p_{34}^2}{p_T^2 p_1\cdot p_2} p_1^{\nu} p_2^{\mu}
-\frac{2 p_1 \cdot p_{34}}{p_T^2 p_1 \cdot p_2}p_2^{\mu} p_{34}^{\nu}
-\frac{2 p_2 \cdot p_{34}}{p_T^2 p_1 \cdot p_2}p_1^{\nu} p_{34}^{\mu}\Big)\nn \\
&+&A_3  \, \Big( g^{\mu \sigma} g^{\nu\rho}
+\frac{g^{\mu \nu} p_1^{\sigma} p_2^{\rho}}{p_1 \cdot p_2}
-\frac{g^{\nu \rho} p_1^{\sigma} p_2^{\mu}}{p_1 \cdot p_2}
-\frac{g^{\mu \sigma} p_1^{\nu} p_2^{\rho}}{p_1 \cdot p_2}\Big) \nn \\
&+&A_4  \, \Big( g^{\mu \rho} g^{\nu\sigma}
+\frac{g^{\mu \nu} p_1^{\rho} p_2^{\sigma}}{p_1 \cdot p_2}
-\frac{g^{\nu \sigma} p_1^{\rho} p_2^{\mu}}{p_1 \cdot p_2}
-\frac{g^{\mu \rho} p_1^{\nu} p_2^{\sigma}}{p_1 \cdot p_2}\Big) \nn \\
&+&A_5  \, \frac{1}{p_1 \cdot p_2} \Big( 
g^{\mu \sigma} p_1^{\rho} p_{34}^{\sigma}
-g^{\mu \rho} p_1^{\sigma} p_{34}^{\nu}
+g^{\nu \sigma} p_2^{\rho} p_{34}^{\mu}
-g^{\nu \rho} p_2^{\sigma} p_{34}^{\mu} \nn \\
&+&\frac{p_2 \cdot p_{34}}{p_1 \cdot p_2}g^{\mu \rho} p_1^{\nu} p_1^{\sigma}
-\frac{p_2 \cdot p_{34}}{p_1 \cdot p_2}g^{\mu \sigma} p_1^{\nu} p_1^{\rho}
+\frac{p_1 \cdot p_{34}}{p_1 \cdot p_2}g^{\nu \rho} p_2^{\mu} p_2^{\sigma}
-\frac{p_1 \cdot p_{34}}{p_1 \cdot p_2}g^{\nu \sigma} p_2^{\mu} p_2^{\rho}\Big)\nn \\
&+&A_6  \, \frac{1}{p_1 \cdot p_2} \Big( 
 g^{\mu \sigma} {p_1^{\rho}   p_{34}^{\nu}}
-g^{\mu \rho}   {p_1^{\sigma} p_{34}^{\nu}}
+g^{\mu \rho}   {p_1^{\nu} p_1^{\sigma}} \frac{p_2 \cdot p_{34}}{p_1 \cdot p_2}
-g^{\mu \sigma} {p_1^{\nu} p_1^{\rho}}  \frac {p_2 \cdot p_{34}}{p_1 \cdot p_2}
\Big) \, .
\eeqn

The six form factors $A_i$ are given by, ($Y=\sud p_T^2= 4 \, p_{34}.p_1 \, p_{34}.p_2 -\sud \stq$)
\beqn
\label{Aiformfactors}
A_1&=&\frac{m^2}{2 \sud} \Big[
            2 \sutqm C_0(\cuxtq) 
          + 2 \sdtqm C_0(\cdxtq) 
          + 2 \sucsm C_0(\cuxcs)
          + 2 \sdcsm C_0(\cdxcs) \nn \\
          &-& 2 Y D_0(\duxtqxd)
          + \sud (\sud-4 m^2) (D_0(\duxtqxd)+D_0(\ddxuxtq)+D_0(\duxdxtq))\Big]\nn \\
A_2 &=& 2 m^2 \Big[ D_0^{d=6}(\duxdxtq)+D_0^{d=6}(\ddxuxtq) +C_0(\cudxtq) 
      +m^2 \Big(D_0(\duxdxtq)+D_0(\ddxuxtq)-D_0(\duxtqxd)\Big) \Big] \nn \\
A_3&=& \frac{1}{2} m^2 \sud \Big[
            D_0(\duxdxtq)
          - D_0(\ddxuxtq)
          - D_0(\duxtqxd)\Big]\nn \\
A_4&=& \frac{1}{2} m^2 \sud \Big[
            D_0(\ddxuxtq)
          - D_0(\duxdxtq)
          - D_0(\duxtqxd)\Big] \\
A_5 &=& \frac{m^2 s}{2 \sdtq \sutq } \Big[ 2 \sutq D_0^{d=6}(\duxdxtq)
      + 2\sdtq D_0^{d=6}(\ddxuxtq) -\sdtq \sutq D_0(\duxtqxd) \nn \\
   &+&4m^2 \Big(\sutq D_0(\duxdxtq)+\sdtq D_0(\ddxuxtq)\Big)+2(\sdtq+\sutq)C_0(\cudxtq)\Big] \nn \\
A_6&=&-\frac{  m^2 \sud}{Y} \Big[         
\sutqm  C_0(\cuxtq) -\sdtqm C_0(\cdxtq) +\sucsm C_0(\cuxcs)-\sdcsm C_0(\cdxcs) \Big] \equiv 0 \nn \; .
\eeqn
In writing these equations we have
introduced the notation,
\beq
s_{ij} = (p_i+p_j)^2 \, ,
s_{ijk} = (p_i+p_j+p_k)^2 \,.
\eeq
We also note the following relations,
\beqn
\sut + \suq\, & \equiv & 2p_1 \cdot p_{34} \equiv \spab{1}.{(3+4)}.{1} \, , \nn \\
\sdt + \sdq\, & \equiv & 2p_2 \cdot p_{34} \equiv \spab{2}.{(3+4)}.{2} \, , \nn \\
\suc + \sus\, & \equiv & 2p_1 \cdot p_{56} \equiv \spab{1}.{(5+6)}.{1} \, , \nn \\
\sdc + \sds\, & \equiv & 2p_2 \cdot p_{56} \equiv \spab{2}.{(5+6)}.{2} \, .
\eeqn
We further introduce the following functions that naturally occur in the coefficients of triangle integrals,
\beqn
\label{deltadefs}
\delta_{ij,kl,mn}  &=& s_{ij} - s_{kl} - s_{mn} \;, \nn \\
\Delta_3 &=& s_{12}^2 + s_{34}^2 + s_{56}^2 
- 2 s_{12}s_{34} - 2 s_{34} s_{56} - 2 s_{56} s_{12} \;.
\eeqn

\begin{table}
\begin{center}
\begin{tabular}{|l|l||l|l||l|l|}
\hline
$D_0^{d=6}{(1)}$     &$D_0^{d=6}(p_1,p_{34},p_2;m,m,m,m)$&$C_0{(1)}$&$C_0(p_1,p_2;m,m,m)$      &$B_0{(1)}$&$B_0(p_{12};m,m)$ \\
$D_0^{d=6}{(2)}$     &$D_0^{d=6}(p_2,p_1,p_{34};m,m,m,m)$&$C_0{(2)}$&$C_0(p_{12},p_{34};m,m,m)$&$B_0{(2)}$&$B_0(p_{34};m,m)$ \\    
$D_0^{d=6}{(3)}$     &$D_0^{d=6}(p_2,p_1,p_{34};m,m,m,m)$&$C_0{(3)}$&$C_0(p_{1},p_{34};m,m,m)$ &$B_0{(3)}$&$B_0(p_{56};m,m)$ \\
$D_0{(1)}$           &$D_0(p_1,p_{34},p_2;m,m,m,m)$      &$C_0{(4)}$&$C_0(p_{2},p_{34};m,m,m)$ &$B_0{(4)}$&$B_0(p_{134};m,m)$\\
$D_0{(2)}$           &$D_0(p_2,p_1,p_{34};m,m,m,m)$      &$C_0{(5)}$&$C_0(p_{1},p_{56};m,m,m)$ &$B_0{(5)}$&$B_0(p_{234};m,m)$\\ 
$D_0{(3)}$           &$D_0(p_2,p_1,p_{34};m,m,m,m)$      &$C_0{(6)}$&$C_0(p_{2},p_{56};m,m,m)$ &          &                  \\
\hline
\end{tabular}
\end{center}
\caption{Definitions of the scalar integrals that appear in the calculation
of the amplitude for continuum production of $gg \to ZZ$ through a loop containing a 
massive particle.}
\label{integraldefns}
\end{table}
The notation for the scalar integrals, $D_0(j),C_0(j)$ is given in Table~\ref{integraldefns} and the
explicit expressions for the six-dimensional boxes are, ($p_1^2=p_2^2=0$), (c.f.\ Eq.~(\ref{sixdim})),
\beqn
D_0^{d=6}(\ddxuxtq)&=&\frac{\sutq}{ 2 Y} \Big[
   \sutqm  C_0(\cuxtq) +\sdcsm C_0(\cdxcs)+\sud C_0(\cuxd) \nn \\
   &+& \Big( \sud-\stq -\scs+2 \frac{\stq \scs}{\sutq} \Big) C_0(\cudxtq) 
   -\Big(\sud \sutq+\frac{4 m^2 Y}{\sutq}\Big) D_0(\ddxuxtq)\Big]  \\
D_0^{d=6}(\duxdxtq)&=&\frac{\sdtq}{ 2 Y} \Big[
   \sdtqm  C_0(\cdxtq) +\sucsm C_0(\cuxcs)+\sud C_0(\cuxd) \nn \\
   &+& \Big( \sud-\stq -\scs+2 \frac{\stq \scs}{\sdtq} \Big) C_0(\cudxtq) 
   -\Big(\sud \sdtq+\frac{4 m^2 Y}{\sdtq}\Big) D_0(\duxdxtq)\Big] \, .
\eeqn
Note that the combination of integrals given in $A_6$ is identically equal to zero,
\beq
 \sutqm  C_0(\cuxtq) -\sdtqm C_0(\cdxtq) +\sucsm C_0(\cuxcs)-\sdcsm C_0(\cdxcs) =0 \;,
\eeq
so that $A_6$ can be dropped from further discussion.
These formula, up to an overall factor, are in agreement with the
result given in ref.~\cite{Zecher:1994kb}. In addition, in the limit $p_{34}^2=p_{56}^2=M_Z^2$,
they are in agreement with the formula of ref.~\cite{Glover:1988rg}.
This concludes our discussion of the tensor $P_{LR}^{\mu\nu\rho\sigma}$.

Contracting with the polarization vectors of the gluons, $\epsilon^{\pm}$, we find
helicity amplitudes for the $(1^-,2^+)$ and $(1^+,2^+)$ polarizations are,
\beqn
\epsilon_\mu^-(p_1) \epsilon_\nu^+(p_2) P_{LR}^{\mu\nu\rho\sigma} &=&\frac{1}{2}\frac{1}{s_{12}^2} \Big[
        -2 g^{\rho \sigma} \frac{\spab{1}.{(3+4)}.{2}}{\spab{2}.{(3+4)}.{1}} s_{12}^2 A_2 
       - \spab{1}.{\gamma^\rho}.{2} \spab{1}.{\gamma^{\sigma}}.{2} s_{12} (A_3+A_4) \nn \\
       &-& \spa{1}.{2} \spbab{2}.{\gamma^\rho}.{\gamma^\sigma}.{2} \spab{1}.{(3+4)}.{2} A_5
        + \spaba{1}.{\gamma^\rho}.{\gamma^\sigma}.{1} \spb{1}.{2} \spab{1}.{(3+4)}.{2}  A_5
\Big] \\
\epsilon_\mu^+(p_1) \epsilon_\nu^+(p_2) P_{LR}^{\mu\nu\rho\sigma}  &=&
 \frac{1}{2} \frac{1}{\spa{1}.{2}^2 s_{12}} \Big[
       2 g^{\rho \sigma} s_{12}^2 A_1
       + \spa{1}.{2} \spbab{2}.{\gamma^\rho}.{\gamma^\sigma}.{1} s_{12} A_3
       + \spa{2}.{1} \spbab{1}.{\gamma^\rho}.{\gamma^\sigma}.{2} s_{12} A_4 \nn \\
       &+& \spa{1}.{2} \spbab{2}.{\gamma^\rho}.{\gamma^\sigma}.{2}  \spab{2}.{(3+4)}.{1} A_5 
       +   \spa{2}.{1} \spbab{1}.{\gamma^\rho}.{\gamma^\sigma}.{1}  \spab{1}.{(3+4)}.{2} A_5
\Big]
\eeqn
The final result for the $LR$ amplitude is obtained 
by saturating the indices $\rho$ and $\sigma$ with our standard left-handed currents for the decay into leptons, 
\beq
      \frac{e^2}{\stq \scs} \spab{3}.{\gamma^\rho}.{4} \spab{5}.{\gamma^\sigma}.{6} \;.
\eeq
Thus the amplitude for the standard polarization of the final state leptons can be written
with an overall factor extracted,
\beq
{\cal A}(1_g^{h_1},2_g^{h_2},3_e^-,4_{\eb}^+,5_\mu^-,6_{\bar{\mu}}^+) = \frac{i g_s^2 e^4 }{4 \pi^2} \delta^{C_1 C_2}  
A(1_g^{h_1},2_g^{h_2},3_e^-,4_{\eb}^+,5_\mu^-,6_{\bar{\mu}}^+)
\eeq
where the reduced amplitudes are given in 
terms of the form factors $A_i$ defined in Eq.~(\ref{Aiformfactors}) by,
\beqn
A(1_g^-,2_g^+,3_e^-,4_{\eb}^+,5_\mu^-,6_{\bar{\mu}}^+) &=&  
 \frac{1}{s_{12} s_{34} s_{56}} \Big[
       \spa{3}.{5} \spb{4}.{6} \frac{\spab{1}.{(3+4)}.{2}}{\spab{2}.{(3+4)}.{1}} s_{12} \, A_2 
       -\spa{1}.{3} \spa{1}.{5} \spb{2}.{4} \spb{2}.{6} \, (A_3+A_4) \nn \\
       &+&  \Big(\frac{\spa{3}.{5} \spb{2}.{4} \spb{6}.{2}}{\spb{1}.{2}} 
                +\frac{\spa{1}.{3} \spa{1}.{5} \spb{4}.{6}}{\spa{1}.{2}} \Big) 
            \spab{1}.{(3+4)}.{2} \, A_5 \Big] \, ,\\
A(1_g^+,2_g^+,3_e^-,4_{\eb}^+,5_\mu^-,6_{\bar{\mu}}^+) &=&
   \frac{1}{s_{12} s_{34} s_{56}} \frac{\spa{3}.{5}}{\spa{1}.{2}} \Big[
         \spb{1}.{6} \spb{4}.{2}  s_{12} \, (A_3+A_1)
        -\spb{2}.{6} \spb{4}.{1}  s_{12} \, (A_4+A_1) \nn \\
       &+& \Big( \spb{2}.{4} \spb{2}.{6} \spab{2}.{(3+4)}.{1} 
                -\spb{1}.{4} \spb{1}.{6} \spab{1}.{(3+4)}.{2}\Big) \, A_5
\Big] \, .
\eeqn

\section{Analytic results for the ${LL}$ amplitude}
\label{app:LL}
There are two independent helicity configurations, $(1^+,2^+)$ and $(1^-,2^+)$,
with the remaining helicities obtained using
the $\rm{flip}$ operation defined in Eq.~(\ref{eq:flipdef}), as before,
\beqn
\label{eq:ALLflip}
A_{LL}(1_g^-,2_g^-,3_{e}^-,4_{\eb}^+,5_{\mu}^-,6_{\mub}^+)&=& {\rm flip} \left\{ A_{LL}(1_g^+,2_g^+,3_{e}^-,4_{\eb}^+,5_{\mu}^-,6_{\mub}^+) \right\} \\
A_{LL}(1_g^+,2_g^-,3_{e}^-,4_{\eb}^+,5_{\mu}^-,6_{\mub}^+)&=& {\rm flip} \left\{ A_{LL}(1_g^-,2_g^+,3_{e}^-,4_{\eb}^+,5_{\mu}^-,6_{\mub}^+) \right\} \, .
\eeqn

The amplitude is described by the expansion in scalar integrals in Eq.~(\ref{Melrose_expansion}).
In this appendix we provide explicit expressions for the coefficients that appear there.
The box and triangle coefficients have the general form,
\beqn
d_i(1^{h_1},2^{h_2}) &=& d_i^{(0)}(1^{h_1},2^{h_2})+m^2 d_i^{(2)}(1^{h_1},2^{h_2})+m^4 d_i^{(4)}(1^{h_1},2^{h_2}) \, , \\
c_i(1^{h_1},2^{h_2}) &=& c_i^{(0)}(1^{h_1},2^{h_2})+m^2 c_i^{(2)}(1^{h_1},2^{h_2})\, ,
\eeqn
while the 6-dimensional box and bubble coefficients, denoted by $d_i^{\, d=6}$ and $b_i^{(0)}$, are independent of the mass.
The latter fact allows the bubble coefficients to be extracted from ref.~\cite{Bern:1997sc}.   
Furthermore the bubble coefficients are constrained by the absence of ultraviolet divergences,
\beq
\label{UVfiniteness}
\sum_{j=1,5} b_j^{(0)}=0 \,.
 \eeq

In the limit $m \to 0$ diagrams develop infra-red poles which must vanish.
Since the amplitude that we are calculating is finite in the $m \to 0$ limit, 
all of the poles in $\epsilon$ must cancel. 
This cancellation of
infra-red poles leads to relations between $m=0$ parts of the triangle and box coefficients.
\begin{eqnarray}
&& \frac{2\, d_1^{(0)}}{(\sutq \sdtq -\stq \scs)}
  +\frac{c_3^{(0)}}{\sutqm}+\frac{c_4^{(0)}}{\sdtqm}-\frac{c_1^{(0)}}{\sud} =0 \;, \\
&& \frac{c_3^{(0)}}{\sutqm}+\frac{c_4^{(0)}}{\sdtqm}
  -\frac{c_5^{(0)}}{\sucsm}-\frac{c_6^{(0)}}{\sdcsm} =0 \;, \\
&& \frac{d_2^{(0)}}{\sud \sutq}-\frac{d_3^{(0)}}{\sud \sdtq}
  +\frac{c_3^{(0)}}{\sutqm}-\frac{c_5^{(0)}}{\sucsm}=0 \;.
\end{eqnarray}
Because of the use of the 6-dimensional box in the $(1^-,2^+)$ amplitude these relations are
trivial in that case.  Nevertheless they provide useful constraints for the $(1^+,2^+)$ coefficients.

\subsection{$(1^+,2^+)$ $d=6$ boxes}
For this helicity combination these coefficients are equal to zero.

\subsection{$(1^+,2^+)$ boxes}
\subsubsection{Box 1:~$D_0(p_1,p_{34},p_2;m,m,m,m)$}
\beqn
      d_1^{(0)}(1^{+},2^{+})&=&\frac{1}{2} \frac{(2 \spa{1}.{3} \spa{2}.{3} \spa{1}.{5} \spa{2}.{5} 
       +\spa{3}.{5}^2 \spa{1}.{2}^2) (\sutq \sdtq-\stq \scs)}
     {\spa{1}.{2}^4 \spa{3}.{4} \spa{5}.{6}}  \\
  d_1^{(2)}(1^{+},2^{+})&=&\Big\{ -\frac{1}{2} \Big[  
     \spa{1}.{2} \spa{1}.{3} \spa{2}.{5} \spa{3}.{4} \spb{4}.{2} \spb{4}.{3} \spb{6}.{1}  
       +\spa{1}.{2} \spa{1}.{3} \spa{1}.{5} \spa{2}.{3} \spb{2}.{1} \spb{4}.{3} \spb{6}.{1}   \nn \\
     &-&  \spa{1}.{2}^2 \spa{3}.{4} \spa{3}.{5} \spb{4}.{2} \spb{4}.{3} \spb{6}.{1}   
       -\spa{1}.{2}\spa{1}.{5} \spa{2}.{3} \spab{5}.{(3+4)}.{2} \spb{4}.{1} \spb{6}.{5}   \nn \\
     &-&  4 \spa{1}.{5}^2 \spa{2}.{3}^2 \spb{2}.{1} \spb{4}.{3} \spb{6}.{5}   
       -\spa{1}.{2}\spa{1}.{3} \spa{2}.{5} \spa{3}.{5} \spb{2}.{1} \spb{4}.{3} \spb{6}.{5}   \nn \\
     &+&  \spa{1}.{2}^2 \spa{3}.{5}^2 \spb{2}.{1} \spb{4}.{3} \spb{6}.{5}\Big]
       \frac{1}{\spa{1}.{2}^3 s_{34} s_{56}}\Big\} + \Big\{ 1 \leftrightarrow 2\Big\}  \\
  d_1^{(4)}(1^+,2^+)&=&\frac{2}{s_{34} s_{56} \spa{1}.{2}^2 \, \spab{1}.{(3+4)}.{2} \spab{2}.{(3+4)}.{1}} \nn \\
       &\times &  \Big(\spb{4}.{2} \spab{2}.{(3+4)}.{1} \spa{1}.{5}
                      -\spb{4}.{1} \spab{1}.{(3+4)}.{2} \spa{2}.{5} \Big) \nn \\
       &\times &  \Big(\spb{6}.{1} \spab{1}.{(3+4)}.{2} \spa{2}.{3} 
                      -\spb{6}.{2} \spab{2}.{(3+4)}.{1} \spa{1}.{3} \Big) 
\eeqn
\subsubsection{Box 2:~$D_0(p_2,p_1,p_{34};m,m,m,m)$}
\beqn
d_2^{(0)}(1^+,2^+)&=&0 \\
d_2^{(2)}(1^+,2^+)&=& \Bigg\{
  \frac{\spb{2}.{1}\spb{4}.{3}
  \big(\spa{1}.{5}\spab{1}.{(3+4)}.{6}{\spab{3}.{(5+6)}.{2}}^2
  -\spa{1}.{3}\spa{5}.{6}\spaba{1}.{(3+4)}.{(5+6)}.{3}{\spb{6}.{2}}^2\big)}
  {2 \, s_{34}s_{56}\spa{1}.{2}{\spab{1}.{(3+4)}.{2}}^2}\Bigg\}
\nn\\
&+& \Bigg\{ (1 \leftrightarrow 2),(3 \leftrightarrow 5),(4 \leftrightarrow 6) \Bigg\} \\
d_2^{(4)}(1^+,2^+)&=&d_1^{(4)}(1^+,2^+)
\eeqn
\subsubsection{Box 3:~$D_0(p_1,p_2,p_{34};m,m,m,m)$}
The results for $d_3$ can be found from the results for $d_2$
by applying the operation $\rm{flip}_2$ defined by,
\beq
\label{eq:flip2def}
{\rm flip}_2: \qquad (3 \leftrightarrow 5), \; (4 \leftrightarrow 6) ,\; (h_3 \leftrightarrow h_5), \; (h_4 \leftrightarrow h_6) \,. 
\eeq
Note that under this operation the helicities of the lepton lines are switched.
\subsection{$(1^+,2^+)$ triangles}
\subsubsection{Triangle 1:~ $C_0(p_1,p_2;m,m,m)$}
\beqn
c_1^{(0)}(1^{+},2^{+})&=&0 \\
c_1^{(2)}(1^{+},2^{+})&=&\frac{1}{s_{34} s_{56}} \frac{\spb{2}.{1}}{\spa {1}.{2}}  \nn \\
     &\times & \Big\{\Big[
              \frac{\spa{1}.{3} \spa{1}.{5} \spb{2}.{4} \spb{2}.{6} (\sutq+\sdtq)}
              {\spab{1}.{(3+4)}.{2}^2}   
            + \frac{(\spa{1}.{3} \spa{5}.{6} \spb{2}.{6} 
                  + \spa{1}.{5} \spa{3}.{4} \spb{2}.{4}) \spb{4}.{6}}{\spab{1}.{(3+4)}.{2}} \Big]   \nn \\
     &+&  \Big[ (1 \leftrightarrow 2),(3 \leftrightarrow 5),(4 \leftrightarrow 6) \Big]\Big\}
\eeqn
\subsubsection{Triangle 2:~$C_0(p_{12},p_{34};m,m,m)$}
\beq
      c_2^{(0)}(1^+,2^+)=0 \, .
\eeq
The mass dependent piece $c_2^{(2)}(1^+,2^+)$ is non-zero.  It is obtained
by exploiting the relation between mass dependent terms in the box and triangle coefficients,
and the rational term~\cite{Badger:2008cm}.  The relation is,
\beq
\label{rattrick}
R(1^{h_1},2^{h_2})
 -\frac{1}{2} \sum_{j=1}^{6} c_j^{(2)}(1^{h_1},2^{h_2}) + \frac{1}{6} \sum_{j=1}^{3} d_j^{(4)}(1^{h_1},2^{h_2}) =0\, .
\eeq

\subsubsection{Triangle 3:~ $C_0(p_1,p_{34};m,m,m)$}
\beqn
c_3^{(0)}(1^{+},2^{+})
   &=&-\frac{1}{2} 
 \frac{(\spa{1}.{5}^2 \spa{2}.{3}^2+\spa{1}.{3}^2 \spa{2}.{5}^2)\sutqm}
{\spa{1}.{2}^4 \spa{3}.{4} \spa{5}.{6}} \\
c_3^{(2)}(1^{+},2^{+}) &=&
     \frac{1}{(s_{34} s_{56})}  \Big[   
     \frac{\spa{1}.{3} \spa{2}.{5} \spab{1}.{(3+4)}.{6} \spb{4}.{1}}  
      {\spa{1}.{2}^3}  
       -\frac{\spa{1}.{3}^2 \spa{2}.{5} \spab{1}.{(3+4)}.{6} \spb{2}.{1} \spb{4}.{3}}  
       {\spa{1}.{2}^3 \spab{1}.{(3+4)}.{2}}   \nn \\
     &+& \frac{\spa{1}.{3} \spab{2}.{(1+3)}.{4} \spab{5}.{(3+4)}.{1} \spb{6}.{1}}
       {\spa{1}.{2}^2 \spab{2}.{(3+4)}.{1}}  
      -\frac{\spa{1}.{3}^2 \spa{1}.{5} \spb{2}.{1} \spb{4}.{3} \spb{6}.{1}}
      {\spa{1}.{2}^2 \spab{1}.{(3+4)}.{2}}  \nn \\
     &+& \frac{2 \spa{1}.{3} \spa{1}.{5} \spa{3}.{4} \spb{4}.{1} \spb{4}.{3} \spb{6}.{1}}
      {\spa{1}.{2}^2 \spab{1}.{(3+4)}.{1}}   
     +\frac{2 \spa{1}.{3} \spab{5}.{(3+4)}.{1} \spb{4}.{1} \spb{6}.{2}}
      {\spa{1}.{2} \spab{1}.{(3+4)}.{1}}   \nn \\
     &-& \frac{\spa{1}.{3} \spa{3}.{5} \spab{1}.{(3+4)}.{1} \spb{4}.{3} \spb{6}.{2}}
      {\spa{1}.{2}^2 \spab{1}.{(3+4)}.{2}} 
     -\frac{\spa{1}.{3}^2 \spa{1}.{5} \spab{1}.{(3+4)}.{1}
      \spab{2}.{(5+6)}.{2} \spb{4}.{3} \spb{6}.{2}}
       {\spa{1}.{2}^3 \spab{1}.{(3+4)}.{2}^2}  \nn \\
   &+&\frac{\spa{1}.{3} \spa{1}.{5} \spa{2}.{3} \spb{2}.{1} \spb{4}.{3} \spb{6}.{2}}
     {\spa{1}.{2}^2 \spab{1}.{(3+4)}.{2}}   
   +\frac{\spa{1}.{3} \spa{1}.{5} \spab{1}.{(3+4)}.{1} \spab{2}.{(3+4)}.{2} \spb{6}.{4}}
     {\spa{1}.{2}^3 \spab{1}.{(3+4)}.{2}}   \nn \\
   &+&\frac{\spa{2}.{3} \spab{5}.{(3+4)}.{1}^2 \spb{4}.{1} \spb{6}.{5}}
     {\spa{1}.{2} \spab{2}.{(3+4)}.{1}^2}\Big] 
\eeqn

\subsubsection{Triangles 4,5 and 6 :~ $C_0(p_{2},p_{34};m,m,m),C_0(p_{1},p_{56};m,m,m)$ and $C_0(p_{2},p_{56};m,m,m)$}
These triangle coefficients can all be obtained from $c_3$ by various symmetry operations.  To that end we define,
\beq
\label{eq:flip1def}
{\rm flip}_1: \qquad (1 \leftrightarrow 2),\; (h_1 \leftrightarrow h_2) \,. 
\eeq
and 
\beq
\label{eq:flip3def}
{\rm flip}_3: \qquad (1 \leftrightarrow 2),\;  (3 \leftrightarrow 5), \; (4 \leftrightarrow 6) ,\;
(h_1 \leftrightarrow h_2) ,\;(h_3 \leftrightarrow h_5), \; (h_4 \leftrightarrow h_6) \,, 
\eeq
and note that ${\rm flip}_3$ is equivalent to applying both ${\rm flip}_1$ and ${\rm flip}_2$ (defined in Eq.~\ref{eq:flip2def}).
Explicitly,
\beqn
c_4(1^+,2^+) &=& {\rm flip}_1 \left\{ c_3(1^+,2^+) \right\} \,, \\
c_5(1^+,2^+) &=& {\rm flip}_2 \left\{ c_3(1^+,2^+) \right\} \,, \\
c_6(1^+,2^+) &=& {\rm flip}_3 \left\{ c_3(1^+,2^+) \right\} \,,
\eeqn

\subsection{$(1^+,2^+)$ bubbles}
Note that the bubble coefficients do not have terms of order $m^2$ in the coefficients.

\subsubsection{Bubble 1 :~$B_0(p_{12};m,m)$}
\beqn 
b_1^{(0)}(1^+,2^+) &=& 0 \, .
\eeqn

\subsubsection{Bubble 2 :~$B_0(p_{34};m,m)$}
\beqn
b_2^{(0)}(1^+,2^+) &=& \frac{\spa{3}.{4}}{\spa{1}.{2}^2 \spa{5}.{6}}   
        \Big[\frac{\spa{1}.{5}^2 \spb{1}.{4}^2}{\sutqm^2}   
        +\frac{\spa{2}.{5}^2 \spb{2}.{4}^2}{\sdtqm^2}\Big]   \nn \\
     &+&  2 \frac{\spa{1}.{5} \spa{2}.{5}}{\spa{1}.{2}^3 \spa{5}.{6}} 
        \Big[\frac{\spa{3}.{1} \spb{4}.{1}}{\sutqm}   
            +\frac{\spa{2}.{3} \spb{4}.{2}}{\sdtqm}\Big]
\eeqn

\subsubsection{Bubble 3 :~$B_0(p_{56};m,m)$}
\beqn
b_3^{(0)}(1^+,2^+)&=& {\rm flip}_2 \left\{ b_2^{(0)}(1^+,2^+) \right\}
\eeqn
\subsubsection{Bubble 4 :~$B_0(p_{134};m,m)$}
\beqn
      b_4^{(0)}(1^+,2^+) &=&-\Bigg[
  \spa{1}.{2}  
      \Big[\frac{\spa{1}.{5}^2 \spa{3}.{4}^2 \spb{1}.{4}^2}{\sutqm^2}   
          +\frac{\spa{2}.{3}^2 \spa{5}.{6}^2 \spb{2}.{6}^2}{\sdcsm^2}\Big] \nn \\
   &+& 2  \spa{1}.{3} \spa{2}.{5}   
     \Big[\frac{\spa{1}.{5} \spa{3}.{4} \spb{1}.{4} }{\sutqm}
         -\frac{\spa{2}.{3} \spa{5}.{6} \spb{2}.{6}}{\sdcsm}   \Big]   \nn \\
     &+& \big(\spa{1}.{3} \spa{2}.{5}+\spa{2}.{3} \spa{1}.{5}\big) \spa{3}.{5}\Bigg]   
     \times   \frac{1}{(\spa{1}.{2}^3 \spa{3}.{4} \spa{5}.{6})}  
\eeqn

\subsubsection{Bubble 5 :~$B_0(p_{234};m,m)$}
\beqn
b_5^{(0)}(1^+,2^+)&=& {\rm flip}_1 \left\{ b_4^{(0)}(1^+,2^+) \right\}
\eeqn

\subsection{$(1^+,2^+)$ rational terms}
\beqn
      R(1^+,2^+)&=&\Bigg[  
        \bigg(\frac{\spa{1}.{5}^2 \spb{1}.{4}^2}{\sutqm}  
             +\frac{\spa{2}.{5}^2 \spb{2}.{4}^2}{\sdtqm}  \bigg)  
       \frac{1}{\spa{5}.{6} \spb{3}.{4}}
        +\bigg(\frac{\spa{3}.{1}^2 \spb{1}.{6}^2}{\sucsm}   
              +\frac{\spa{3}.{2}^2 \spb{2}.{6}^2}{\sdcsm}\bigg)  
        \frac{1}{\spb{5}.{6} \spa{3}.{4}}   \nn \\
     &+&  \frac{\spb{4}.{6}^2}{\spb{3}.{4} \spb{5}.{6}}
         -\frac{\spa{3}.{5}^2}{\spa{3}.{4} \spa{5}.{6}}  \Bigg]\frac{1}{\spa{1}.{2}^2}   
\eeqn

\subsection{$(1^-,2^+)$ $d=6$ boxes}

\subsubsection{Box 2:~$D_0^{d=6}(p_2,p_1,p_{34};m,m,m,m)$}

\beqn
d_2^{d=6}(1^-,2^+)&=& 
      \frac{-1}{\spb{3}.{4} \spa{5}.{6} \sutq}
   \frac{\spab{1}.{(3+4)}.{2}}{\spab{2}.{(3+4)}.{1}^3} 
       \Big[\spab{2}.{(1+3)}.{4}^2 \spab{5}.{(3+4)}.{1}^2
       +\sutq^2 \spa{2}.{5}^2 \spb{1}.{4}^2\Big]
\eeqn

\subsubsection{Box 3:~$D_0^{d=6}(p_1,p_2,p_{34};m,m,m,m)$}

\beq
d_3^{d=6}(1^+,2^-)= {\rm flip}_2 \left\{ d_2^{d=6}(1^-,2^+) \right\}
\eeq
Note that the standard helicity choice, $(1^-,2^+)$, can be recovered by applying the ${\rm flip}$ operation
defined in Eq.~(\ref{eq:flipdef}).

\subsection{$(1^-,2^+)$ boxes}

\subsubsection{Box 1:~$D_0(p_1,p_{34},p_2;m,m,m,m)$}
\beqn
      {d}_1^{(0)}(1^{-},2^{+})&=&0 \\
      {d}_1^{(2)}(1^{-},2^{+})&=&\Big\{\frac{1}{2} \Big[-   
       \frac{\spa{5}.{6} \spab{1}.{(3+4)}.{2} \spab{3}.{(2+4)}.{1}   
        \spb{4}.{1} \spb{6}.{2}^2}
         {\spab{2}.{(3+4)}.{1} \spb{2}.{1}^2}   \nn \\
     &+&  \frac{\spa{1}.{5}^2 \spa{2}.{3} 
       \spab{1}.{(3+4)}.{2} \spab{2}.{(1+3)}.{4} \spb{6}.{5}}  
        {\spa{1}.{2}^2 \spab{2}.{(3+4)}.{1}}\Big]\frac{1}{s_{34} s_{56}} \Big\}
       +\Big\{ 3 \leftrightarrow 5, 4 \leftrightarrow 6, \Big\}   \\
      {d}_1^{(4)}(1^{-},2^{+}) &=&  
   \frac{\spab{1}.{(3+4)}.{2} \spa{2}.{1}}{\spba{1}.{(3+4)}.{2} \spb{2}.{1} } \, d_1^{(4)}(1^+,2^+) \nn \\
     &=&\frac{2}{ s_{34} s_{56}   \spa{1}.{2} \spb{1}.{2} \, \spab{2}.{(3+4)}.{1}^2 } \nn \\
     &\times&   \Big(\spb{4}.{2} \spab{2}.{(3+4)}.{1}\spa{1}.{5}
                    -\spb{4}.{1} \spab{1}.{(3+4)}.{2}\spa{2}.{5}   \Big)\nn \\
     &\times&   \Big(\spb{6}.{1} \spab{1}.{(3+4)}.{2} \spa{2}.{3} 
                    -\spb{6}.{2} \spab{2}.{(3+4)}.{1} \spa{1}.{3}  \Big)
\eeqn

\subsubsection{Box 2:~$D_0(p_2,p_1,p_{34};m,m,m,m)$}
\beqn
{d}_2^{(2)}(1^-,2^+)&=& \frac{1}{s_{34} s_{56}} \Bigg\{
    \frac{\spa{2}.{5} \spb{1}.{4} }{\spab{2}.{(3+4)}.{1}^2} \Big[   
        -\frac{1}{2} \sutq^2 \big(\spa{1}.{3} \spb{2}.{6}+\spa{1}.{5} \spa{2}.{3}   
             \frac{\spb{1}.{6} \spb{2}.{4}}{\spa{2}.{5} \spb{1}.{4}} \big)   \nn \\
     &-&   \frac{1}{2} \sutq \Big(
                        \spa{1}.{5} \spa{3}.{4} \spb{2}.{4} \spb{5}.{6}   
                        -\spa{3}.{5} \spb{5}.{6} \spab{1}.{(3+4)}.{2}   \nn \\
     &+&                    2 \spa{3}.{4} \spb{4}.{6} \spab{1}.{(3+4)}.{2}   
                         +\spab{1}.{(3+4)}.{6} \spab{3}.{(1+4)}.{2}\Big)   \nn \\
     &+&   \frac{1}{2} \spa{3}.{4} \spb{2}.{6} \spab{2}.{(5+6)}.{4} \spab{1}.{(3+4)}.{2}   
         -2 \spa{3}.{4} \spb{5}.{6} \spab{1}.{(3+4)}.{2} \spab{5}.{(1+3)}.{4}\Big]   \nn \\
     &+&  \frac{1}{\spab{2}.{(3+4)}.{1}} \Big[
        -\frac{1}{2} \sutq \spa{1}.{3} \spb{2}.{6} \spab{5}.{(1+3)}.{4}   
        -\frac{2}{\sutq} \spa{3}.{4} \spb{5}.{6} \spab{1}.{(3+4)}.{2} \spab{5}.{(1+3)}.{4}^2   \nn \\
     &-&   \frac{1}{2} \spa{1}.{3} \spa{1}.{5} \spb{2}.{4} \spb{5}.{6} \spab{5}.{(3+4)}.{1} 
       -\frac{1}{2} \spa{1}.{5} \spa{3}.{4} \spb{2}.{4} \spb{2}.{6} \spab{2}.{(5+6)}.{4}   \nn \\
     &+&   \frac{1}{2} \spab{1}.{(3+4)}.{2} \spab{5}.{(1+3)}.{4}  (\spa{2}.{3} \spb{2}.{6}+\spa{3}.{5} \spb{5}.{6}   
           -2 \spa{3}.{4} \spb{4}.{6})\Big] \Bigg\} \nn \\
  {d}_2^{(4)}(1^-,2^+)&=&{d}_1^{(4)}(1^-,2^+)
\eeqn

\subsubsection{Box 3:~$D_0(p_1,p_2,p_{34};m,m,m,m)$}
\beq
d_3(1^+,2^-)= {\rm flip}_2 \left\{ d_2(1^-,2^+) \right\}
\eeq
Note that the standard helicity choice can be recovered by applying the ${\rm flip}$ operation
defined in Eq.~(\ref{eq:flipdef}).

\subsection{$(1^-,2^+)$ triangles}

\subsubsection{Triangle 1:~ $C_0(p_1,p_2;m,m,m)$}
\beqn
{c}_1^{(0)}(1^{-},2^{+}) &=& 0 \\
{c}_1^{(2)}(1^{-},2^{+}) &=& \frac{1}{\stq\scs}\Big[-2 \frac{\spab{1}.{(3+4)}.{2}}{\spab{2}.{(3+4)}.{1}^3}
          (\sutq+\sdtq) \spa{2}.{3} \spa{2}.{5} \spb{1}.{4} \spb{1}.{6}  \nn\\
     &+&  \frac{1}{\spab{2}.{(3+4)}.{1}^2} \Big(   
        (\sutq+\sdtq) \big(\spa{1}.{3} \spa{2}.{5} \spb{1}.{4} \spb{2}.{6}
                          +\spa{1}.{5} \spa{2}.{3} \spb{1}.{6} \spb{2}.{4}\big)   \nn \\
     &+&     \spa{2}.{3} \spb{1}.{6} \spab{5}.{(1+6)}.{4}\spab{1}.{(3+4)}.{2}   
            -\spa{2}.{5} \spb{1}.{4} \spab{3}.{(1+4)}.{6} \spab{1}.{(3+4)}.{2}\Big)   \nn \\
     &-&  \spa{3}.{5} \spb{4}.{6} \frac{\spab{1}.{(3+4)}.{2}}{\spab{2}.{(3+4)}.{1}}  \Big]
\eeqn
\subsubsection{Triangle 2:~$C_0(p_{12},p_{34};m,m,m)$}
\beqn
c_2^{(0)}(1^{-},2^{+}) &=&
          6 \frac{1}{\Delta_3^2} s_{12} \frac{\spab{1}.{(3+4)}.{2}}{\spab{2}.{(3+4)}.{1}}
            \spab{3}.{(1+2)}.{4} \spab{5}.{(1+2)}.{6} \delta_{12,34,56}   \nn \\
     &+&  2 \frac{1}{\Delta_3 \spab{2}.{(3+4)}.{1}} \Big[ \spa{2}.{3} \spa{2}.{5} \spb{1}.{4} \spb{1}.{6} 
          \frac{\spab{1}.{(3+4)}.{2}^2}{\spab{2}.{(3+4)}.{1}}   
        +\spa{1}.{3} \spa{1}.{5} \spb{2}.{4} \spb{2}.{6} \spab{2}.{(3+4)}.{1} \nn \\
     &+&   (\sutq-\sdtq)   \frac{\spab{1}.{(3+4)}.{2}}{\spab{2}.{(3+4)}.{1}^2}  
           (\spa{2}.{5}^2 \spa{3}.{4} \spb{1}.{4}^2 \spb{5}.{6}   
                       -\spa{2}.{3}^2 \spa{5}.{6} \spb{1}.{6}^2 \spb{3}.{4})   \nn \\
     &+&   (\sutq-\sdtq) \frac{\spab{1}.{(3+4)}.{2}}{\spab{2}.{(3+4)}.{1}}  
          \big(\spa{2}.{3} \spb{1}.{6} \spab{5}.{(1+3)}.{4}
              +\spa{2}.{5} \spb{1}.{4} \spab{3}.{(1+5)}.{6}\big)   \nn \\
     &-&  3 \spab{1}.{(3+4)}.{2} \big(\spa{1}.{3} \spa{2}.{5} \spb{1}.{4} \spb{2}.{6}
                               +\spa{1}.{5} \spa{2}.{3} \spb{1}.{6} \spb{2}.{4}\big) \Big] \nn \\
     &-&  \frac{\spa{1}.{3} \spb{2}.{6} \spab{5}.{(1+3)}.{4}}{\sutq \spab{2}.{(3+4)}.{1}}  
         +\frac{\spa{1}.{5} \spb{2}.{4} \spab{3}.{(1+5)}.{6}}{\sdtq  \spab{2}.{(3+4)}.{1}}
\eeqn
The kinematic quantities $\Delta_3$ and $\delta_{12,34,56}$ are defined in Eq.~(\ref{deltadefs}).
The coefficient 
$c_2^{(2)}(1^{-},2^{+})$ is again obtained by exploiting the relation between mass-dependent coefficients of
boxes and triangles and the total rational contribution, c.f. Eq.~(\ref{rattrick}).
 
\subsubsection{Triangle 3:~ $C_0(p_1,p_{34};m,m,m)$}
\beqn
c_3^{(0)}(1^{-},2^{+}) &=&  0 \\ 
c_3^{(2)}(1^{-},2^{+}) &=&  \frac{1}{s_{12} s_{34} s_{56}} \Bigg\{  
      -2 \frac{\spab{1}.{(3+4)}.{2}}{\spab{2}.{(3+4)}.{1}^3} \sutq   
         \spb{1}.{4}^2 \spb{1}.{6} \spa{2}.{5} \spa{4}.{3} \spa{2}.{1}   \nn \\
     &+&  \frac{\spb{1}.{4}}{\spab{2}.{(3+4)}.{1}^2} \Big[ 
         \spab{2}.{(5+6)}.{2} \spa{1}.{3} \spab{5}.{(3+4)}.{1} \spab{1}.{(3+4)}.{6}   \nn \\
     &+&   \spab{2}.{(5+6)}.{2} \spa{1}.{3} s_{34} \spb{1}.{6} \spa{1}.{5}   
        +\spab{1}.{(3+4)}.{2}^2 \frac{\spb{1}.{4} \spb{1}.{6} \spa{2}.{5}\spa{3}.{4}}{\spb{1}.{2}} \nn \\
     &+&   \spb{1}.{4} \spb{2}.{6} \spa{1}.{2}^2 \spa{3}.{4} \spab{5}.{(3+4)}.{1}
     \frac{\spab{1}.{(3+4)}.{2}}{\spab{1}.{(3+4)}.{1}}   \nn \\
     &+&   2 \sutq \spa{3}.{4} \spb{1}.{4} \spb{1}.{6} \spa{1}.{5} \spa{1}.{2}
          \frac{\spab{1}.{(3+4)}.{2}}{\spab{1}.{(3+4)}.{1}}\Big]   \nn \\
     &+&  \frac{\spb{1}.{4}}{\spab{2}.{(3+4)}.{1}} \Big[ 
          (\sdtq+\stq) \spb{2}.{6} \spa{1}.{3} \spa{1}.{5}   \nn \\
     &+&   \spb{1}.{6} \spa{1}.{3} \spa{1}.{5} \spab{1}.{(3+4)}.{2}  
        -\frac{\spb{1}.{4} \spb{1}.{6} \spa{1}.{5} \spa{3}.{4} \spab{1}.{(3+4)}.{2}^2}
          {\spb{1}.{2} \spab{1}.{(3+4)}.{1}}\nn \\
     &-&   2 \frac{\spa{3}.{4} \spb{3}.{4} \spb{1}.{2} \spa{1}.{3} \spa{1}.{5}^2
         \spb{6}.{5}}{\spab{1}.{(3+4)}.{1}}\Big]
     +  \frac{\spb{1}.{2} \spb{3}.{4} \spa{1}.{3}^2 \spa{1}.{5}^2 \spb{6}.{5}}
         {\spa{1}.{2} \spab{1}.{(3+4)}.{1}} \Bigg\}  
\eeqn

\subsubsection{Triangles 4,5 and 6 :~ $C_0(p_{2},p_{34};m,m,m),C_0(p_{1},p_{56};m,m,m)$ and $C_0(p_{2},p_{56};m,m,m)$}
These coefficients may be obtained using symmetries as follows,
\beqn
c_4(1^+,2^-) &=& {\rm flip}_1 \left\{ c_3(1^-,2^+) \right\} \,, \\
c_5(1^-,2^+) &=& {\rm flip}_2 \left\{ c_3(1^-,2^+) \right\} \,, \\
c_6(1^+,2^-) &=& {\rm flip}_3 \left\{ c_3(1^-,2^+) \right\} \,,
\eeqn
Note that the standard helicity choice for $c_4$ and $c_6$ can be recovered by applying the ${\rm flip}$ operation
defined in Eq.~(\ref{eq:flipdef}).

\subsection{$(1^-,2^+)$ bubbles}
\subsubsection{Bubble 1 :~$B_0(p_{12};m,m)$}
The coefficients for this bubble are obtained by exploiting the cancellation of ultraviolet poles
expressed in Eq.~(\ref{UVfiniteness}),
\beq
b_1^{(0)}(1^-,2^+)=-b_2^{(0)}(1^-,2^+)-b_3^{(0)}(1^-,2^+)-b_4^{(0)}(1^-,2^+)-b_5^{(0)}(1^-,2^+) \,.
\eeq

\subsubsection{Bubble 2 :~$B_0(p_{34};m,m)$}
\beqn
b_2^{(0)}(1^-,2^+)= 
     &-& \frac{\spa{4}.{3} \spa{1}.{5}^2 \spb{1}.{4}^2 s_{34}}
        {\sutqm^2 \spa{6}.{5} \spab{2}.{(3+4)}.{1}^2}  
        -\frac{\spb{4}.{3} \spb{2}.{6}^2 \spa{2}.{3}^2 s_{34}}
               {\sdtqm^2 \spb{6}.{5} \spab{2}.{(3+4)}.{1}^2} \nn \\
     &-& 2 \frac{\spa{3}.{2} \spb{1}.{4} s_{34}}{s_{56} \spab{2}.{(3+4)}.{1}^3 \sutqm \sdtqm}
         \Big[\spa{5}.{6} \spb{2}.{6} \spab{1}.{(3+4)}.{6} \spab{2}.{(3+4)}.{1} \nn \\
    &+&  s_{34} \left(\spa{2}.{5} \spb{1}.{6} \spab{1}.{(3+4)}.{2}
                            -\spa{1}.{5} \spb{2}.{6} \spab{2}.{(3+4)}.{1} \right)   
     - \spa{1}.{5} \spb{5}.{6} \spab{5}.{(3+4)}.{2} \spab{2}.{(3+4)}.{1} \Big]   \nn \\
     &+&  6 \frac{1}{\Delta_3^2}   
       \spab{3}.{(1+2)}.{4} \spab{5}.{(1+2)}.{6} \delta_{56,12,34}
       \frac{\spab{1}.{(3+4)}.{2}}{\spab{2}.{(3+4)}.{1}}   \nn \\
     &+&   \frac{1}{\Delta_3}  \Big[  
          2 \frac{(s_{134}-s_{234}) \spab{1}.{(3+4)}.{2}}{s_{56} \spab{2}.{(3+4)}.{1}} \Big(
           \frac{s_{234} \spb{1}.{4} \spb{1}.{6} \spa{2}.{3} \spa{2}.{5}}{\spab{2}.{(3+4)}.{1}^2}
          +\frac{s_{234} \spb{1}.{4} \spa{1}.{5} \spab{3}.{(2+5)}.{6}}{\spab{1}.{(3+4)}.{2} \spab{2}.{(3+4)}.{1}}   \nn \\
     &+&   \frac{\spb{4}.{6} \spa{2}.{5} \spab{3}.{(2+4)}.{1}}{\spab{2}.{(3+4)}.{1}}
          -\frac{s_{134} \spa{2}.{3} \spa{2}.{5} \spb{1}.{4} \spb{1}.{6}}{\spab{2}.{(3+4)}.{1}^2}
          -\frac{s_{134} \spa{2}.{3} \spb{2}.{6} \spab{5}.{(1+6)}.{4}}{\spab{1}.{(3+4)}.{2} \spab{2}.{(3+4)}.{1}}   \nn \\
     &-&   \frac{\spa{3}.{5} \spb{1}.{6} \spab{2}.{(1+3)}.{4}}{\spab{2}.{(3+4)}.{1}} \Big)    \nn \\
     &-& 8 \frac{\spa{1}.{5} \spa{2}.{5} \spb{1}.{4}^2 \spa{3}.{4} \spab{1}.{(3+4)}.{2}} 
          {\spa{5}.{6} \spab{2}.{(3+4)}.{1}^2}   
        -8 \frac{\spa{2}.{3}^2 \spb{3}.{4} \spb{2}.{6} \spb{1}.{6} \spab{1}.{(3+4)}.{2}}
             {\spb{5}.{6} \spab{2}.{(3+4)}.{1}^2}    \nn \\
     &+& 2 \frac{\spa{3}.{4} \spa{1}.{5} \spb{4}.{1} \spb{4}.{6} (\sdtq-\sutq)}
         {\spab{2}.{(3+4)}.{1}^2}   
         +2 \frac{\spa{3}.{2} \spa{3}.{5} \spb{3}.{4} \spb{2}.{6} (\sdtq-\sutq)}
          {\spab{2}.{(3+4)}.{1}^2}   \nn \\
     &+&  2 \frac{\spa{3}.{5} \spb{4}.{6} s_{34} \delta_{34,12,56}}
          {\spab{2}.{(3+4)}.{1}^2} 
       -\frac{\spa{3}.{5}^2 \spb{3}.{4} \delta_{34,12,56} \sdtq}
          {\spa{6}.{5} \spab{2}.{(3+4)}.{1}^2}   \nn \\
     &-&  \frac{\spa{3}.{4} \spb{4}.{6}^2 \delta_{34,12,56} \sutq}
       {\spb{6}.{5} \spab{2}.{(3+4)}.{1}^2} \
       +\frac{\spa{3}.{5}^2 \spb{3}.{4} \spab{1}.{(3+4)}.{2}}
          {\spa{6}.{5} \spab{2}.{(3+4)}.{1}}   \nn \\
     &+&  \frac{\spa{3}.{4} \spb{4}.{6}^2 \spab{1}.{(3+4)}.{2}}
         {\spb{6}.{5} \spab{2}.{(3+4)}.{1}}
       -2 \frac{\spa{3}.{5} \spb{4}.{6} \spb{6}.{5} \spab{1}.{(3+4)}.{2}}
          {\spb{6}.{5} \spab{2}.{(3+4)}.{1}} \Big]
\eeqn
The kinematic quantities $\Delta_3$ and $\delta_{ij,kl,mn}$ are defined in Eq.~(\ref{deltadefs}).
\subsubsection{Bubble 3 :~$B_0(p_{56};m,m)$}
\beqn
b_3^{(0)}(1^-,2^+)&=& {\rm flip}_2 \left\{ b_2^{(0)}(1^-,2^+) \right\} \,.
\eeqn

\subsubsection{Bubble 4 :~$B_0(p_{134};m,m)$}

\beqn
b_4^{(0)}(1^-,2^+)&=&
      2 \frac{\spab{1}.{(3+4)}.{2}}{\spab{2}.{(3+4)}.{1}^3} 
           \frac{\spa{2}.{5}^2 \spb{1}.{4} \spbb{4}.{(5+6)(3+4)}.{1}}{\spa{1}.{2} \spa{5}.{6} \spb{1}.{2} \spb{3}.{4}}   \nn \\
     &+& \frac{1}{\spab{2}.{(3+4)}.{1}^2}   \Bigg[
            \frac{s_{34} \spa{1}.{5}^2 \spa{3}.{4} \spb{1}.{4}^2}{\spa{5}.{6} \spab{1}.{(3+4)}.{1}^2}  
          +   \frac{s_{56} \spa{2}.{5}^2 \spb{2}.{4}^2 \spb{5}.{6}}{\spb{3}.{4} \spab{2}.{(5+6)}.{2}^2} 
          - 2 \frac{s_{34} \spa{1}.{5}^2 \spa{2}.{3} \spb{1}.{4}}{\spa{1}.{2} \spa{5}.{6} \spab{1}.{(3+4)}.{1}}   \nn \\
     &-&    2 \frac{s_{56} \spa{2}.{5} \spb{1}.{6} \spb{2}.{4}^2}{\spb{1}.{2} \spb{3}.{4} \spab{2}.{(5+6)}.{2}}
          + 2 \frac{\spa{1}.{5} \spa{2}.{5} \spb{1}.{4} \spbb{2}.{(3+4)(5+6)}.{4}}{\spa{1}.{2} \spb{1}.{2} \spb{3}.{4} \spa{5}.{6}}
          -   \frac{\spab{5}.{(2+3)}.{4}^2}{\spa{5}.{6} \spb{3}.{4}}  \Bigg]   \nn \\
\eeqn

\subsubsection{Bubble 5 :~$B_0(p_{234};m,m)$}
\beqn
b_5^{(0)}(1^+,2^-)&=& {\rm flip}_1 \left\{ b_4^{(0)}(1^-,2^+) \right\}
\eeqn
Note that the standard helicity choice can be recovered by applying the ${\rm flip}$ operation
defined in Eq.~(\ref{eq:flipdef}).

\subsection{$(1^-,2^+)$ rational terms}
\beqn
      R(1^-,2^+)&=& 
         \Bigg[ \frac{\spa{2}.{3}^2 \spb{2}.{6}^2 \spb{3}.{4}} {\spb{5}.{6} \sdtqm}  
       + \frac{\spa{1}.{5}^2 \spb{1}.{4}^2 \spa{3}.{4}} {\spa{5}.{6} \sutqm}  
       + \frac{\spa{2}.{5}^2 \spb{2}.{4}^2 \spb{5}.{6}} {\spb{3}.{4} \sdcsm}  
        +\frac{\spa{1}.{3}^2 \spb{1}.{6}^2 \spa{5}.{6}} {\spa{3}.{4} \sucsm}   \nn \\
       &-&\frac{\spab{3}.{(1+4)}.{6}^2}{\spa{3}.{4} \spb{5}.{6}}   
         -\frac{\spab{5}.{(2+3)}.{4}^2}{\spb{3}.{4} \spa{5}.{6}}   
        -2 \spa{3}.{5} \spb{4}.{6}\Bigg] \frac{1}{\spab{2}.{(3+4)}.{1}^2}   \nn \\
     &-&  \frac{1}{\Delta_3} \frac{\spab{1}.{(3+4)}.{2}}{\spab{2}.{(3+4)}.{1}} \Bigg(
         4 \spa{3}.{5} \spb{4}.{6} 
       +(s_{12}-s_{34}-s_{56})   
        \big(\frac{\spa{3}.{5}^2}{\spa{3}.{4} \spa{5}.{6}} 
            +\frac{\spb{4}.{6}^2}{\spb{3}.{4} \spb{5}.{6}}\big)\Bigg) 
\eeqn

\section{Numerical results for coefficients}

In this Appendix we present numerical values for the coefficients computed
in Appendix \ref{app:LR},\ref{app:LL}, evaluated at a particular phase space point.  
The chosen
point corresponds to,
\begin{equation}
\begin{array}{rrrrr}
  p_1 &= ( -3.000000000000  ,&  2.121320343560 ,&   1.060660171780 ,&   1.837117307087 ) \\
  p_2 &= ( -3.000000000000  ,& -2.121320343560 ,&  -1.060660171780 ,&  -1.837117307087 ) \\
  p_3 &= (  0.857142857143  ,& -0.315789473684 ,&   0.796850604481 ,&   0.000000000000 ) \\
  p_4 &= (  2.000000000000  ,&  2.000000000000 ,&   0.000000000000 ,&   0.000000000000 ) \\
  p_5 &= (  1.000000000000  ,& -0.184210526316 ,&   0.464829519280 ,&   0.866025403784 ) \\
  p_6 &= (  2.142857142857  ,& -1.500000000000 ,&  -1.261680123761 ,&  -0.866025403784 )
\label{eq:PSpoint}
\end{array}
\end{equation}
Note that all momenta are massless, $p_i^2=0$ and momentum conservation is represented by,
$p_1 + p_2 + p_3 + p_4 + p_5 + p_6 = 0$ so that the energies of $p_1$ and $p_2$ are negative. 
Results for the coefficients appearing in the amplitudes $A(1^+,2^+,3^-,4^+,5^-,6^+)$
and  $A(1^-,2^+,3^-,4^+,5^-,6^+)$ are shown in Table~\ref{table:numerical}.

\begin{table}
\begin{center}
\input{ALLcoefftable.tex}
\caption{Numerical values of coefficients appearing in the amplitudes $A(1^+,2^+,3^-,4^+,5^-,6^+)$
and  $A(1^-,2^+,3^-,4^+,5^-,6^+)$, evaluated at the phase space point given in
Eq.~(\protect\ref{eq:PSpoint}).}
\label{table:numerical}
\end{center}
\end{table}
In summary we give the value of the reduced matrix elements at our standard point 
Eq.~(\ref{eq:PSpoint}) and for a quark of mass $m=0.4255266775$ running in the loop.
We find for the $LR$ combination,
\beqn
A(1^+,2^+,3^-,4^+,5^-,6^+)&=& -0.3327734872\times 10^{-1}   +0.5996051030\times 10^{-2} \, i \nn \\
A(1^-,2^+,3^-,4^+,5^-,6^+)&=& +0.1157034544   -0.7783407466\times 10^{-1} \, i  
\eeqn
We find for the $LL$ combination using the coefficients given in Eq.~(\ref{table:numerical}),
\beqn
A(1^+,2^+,3^-,4^+,5^-,6^+)&=& -0.2809004251\times 10^{-1}   +0.1111561241 \, i \nn \\
A(1^-,2^+,3^-,4^+,5^-,6^+)&=& -0.1213182997\times 10^{-1}   -0.2215976019\times 10^{-1} \, i
\eeqn

Spinor products are defined by,
\begin{eqnarray} 
\spa{p}.{q} &=& \sqrt{p^- q^+} e^{i\varphi_{p}}
              - \sqrt{p^+ q^-} e^{i\varphi_{q}}, 
\nonumber \\ 
\spb{p}.{q} &=& \sqrt{p^+ q^-} e^{-i\varphi_{q}}
 -\sqrt{p^- q^+} e^{-i\varphi_{p}}
\end{eqnarray} 
where,
\beq 
e^{\pm i\varphi_p}\ \equiv\ 
  \frac{ p^1 \pm ip^2 }{ \sqrt{(p^1)^2+(p^2)^2} }
\ =\  \frac{ p^1 \pm ip^2 }{ \sqrt{p^+p^-} }\ ,
\qquad p^\pm\ =\ p^0 \pm p^3.  
\eeq

\bibliography{ZZ_prd}

\end{document}

%% file: macro.tex
\def\cM{\mathcal{M}}
\def\prop#1{{\cal P}_{#1}}
\def\sutqm{(s_{13}+s_{14})}
\def\sdtqm{(s_{23}+s_{24})}
\def\sucsm{(s_{15}+s_{16})}
\def\sdcsm{(s_{25}+s_{26})}
\def\sutq{{s}_{134}}
\def\sdtq{{s}_{234}}
\def\sucs{{s}_{156}}
\def\sdcs{{s}_{256}}
\def\sud{{s}_{12}}
\def\stq{{s}_{34}}
\def\scs{{s}_{56}}
\def\sut{{s}_{13}}
\def\suq{{s}_{14}}
\def\sdt{{s}_{23}}
\def\sdq{{s}_{24}}
\def\suc{{s}_{15}}
\def\sus{{s}_{16}}
\def\sdc{{s}_{25}}
\def\sds{{s}_{26}}
\def\duxtqxd{1}
\def\ddxuxtq{2}
\def\duxdxtq{3}
\def\cuxd{1}
\def\cudxtq{2}
\def\cuxtq{3}
\def\cdxtq{4}
\def\cuxcs{5}
\def\cdxcs{6}
\def\cD{\cal D}
\def\cP{\cal P}
\def\cg{c_\Gamma}
\def\gW{g_W}
\def\rG{r_\Gamma}
\def\jb{i_{b}}
\def\jt{i_{t}}
\def\tb{\bar{t}}
\def\eb{\bar{e}}
\def\mub{\bar{\mu}}
\def\qb{\bar{q}}
\def\spaa#1.#2.#3{\langle\mskip-1mu{#1}|#2|{#3}\mskip-1mu\rangle}
\def\spbb#1.#2.#3{[\mskip-1mu{#1}|#2|{#3}\mskip-1mu]}
\def\spa#1.#2{\left\langle#1\,#2\right\rangle}
\def\spb#1.#2{\left[#1\,#2\right]}
\def\spab#1.#2.#3{\left\langle#1|#2|#3\right]}
\def\spba#1.#2.#3{\left[#1|#2|#3\right\rangle}
\def\spbab#1.#2.#3.#4{[\mskip-1mu{#1}
                  | #2  #3 | {#4}\mskip-1mu]}
\def\spaba#1.#2.#3.#4{\langle\mskip-1mu{#1}
                  | #2  #3 | {#4}\mskip-1mu\rangle}
\def\Wzb{\bar{W}_0}
\def\Pzb{\bar{P}_0}
\def\Pthreeb{\bar{P}_3}
\def\P3b{\bar{P}_3}
\def\Ypb{\bar{Y}_p}
\def\Ypz{{Y}_p(z)}
\def\Ywb{\bar{Y}_w}
\def\Ppb{\bar{P}_+}
\def\Pmb{\bar{P}_-}
\def\Wpb{\bar{W}_+}
\def\Wmb{\bar{W}_-}
\def\cO{{\cal O}}
\def\li{{\rm Li_2}}
\def\g0{\gamma_0}
\def\gp{\gamma^{+}}
\def\gm{\gamma^{-}}
\def\lp{\gamma^{+}}
\def\lm{\gamma^{-}}
\def\xp{x_{+}}
\def\xm{x_{-}}
\def\bentarrow{\:\raisebox{1.3ex}{\rlap{$\vert$}}\!\rightarrow}                 
\def\dkp#1#2#3#4{
\begin{array}{r c l}
#1 & \rightarrow & #2#3 \\
 & & \phantom{\; #2}\bentarrow #4
\end{array}}                                                                    
\def\bothdk#1#2#3#4#5{
\begin{array}{r c l}                                                            
#1 & \rightarrow & #2#3 \\
 & & \:\raisebox{1.3ex}{\rlap{$\vert$}}\raisebox{-0.5ex}{$\vert$} 
\phantom{#2}\!\bentarrow #4 \\
 & & \bentarrow #5                                                              
\end{array}                                                                     
}                                                                               
\newcommand{\kirill}{\colour{red}}
\newcommand{\comment}[1]{{\bf [#1]}}
\newcommand{\beq}{\begin{equation}}
\newcommand{\eeq}{\end{equation}}
\newcommand{\beqn}{\begin{eqnarray}}
\newcommand{\eeqn}{\end{eqnarray}}
\newcommand{\bi}[1]{\bibitem{#1}}
\newcommand{\fr}[2]{\frac{#1}{#2}}
\newcommand{\non}{\nonumber}
\newcommand{\nn}{\nonumber}
\newcommand{\Et}{E_t}
\newcommand{\Pt}{P_t}
\newcommand{\pt}{p_t}
\newcommand{\pb}{p_b}
\newcommand{\pw}{p_W}
\newcommand{\pg}{p_g}
\newcommand\tpW        {{\tilde p}_W}
\newcommand\tpb        {\tilde p_b}
\newcommand{\ar}{\mbox{$\rightarrow$}}
\def\ra{\rightarrow}

\newcommand{\slsh}{\rlap{$\;\!\!\not$}}     
\def\amuh{a_\mu^{{\mathrm had}}}
\def\vec#1{{\mbox{\boldmath$#1$}}}
\def\ket#1{\vert #1 \rangle}
\def\bra#1{\langle #1 \vert}
\newcommand{\as}{\alpha_S}
\newcommand{\p}{\mbox{$\vec{p}$}}
\newcommand{\q}{\mbox{$\vec{q}$}}
\newcommand{\pp}{\mbox{$\vec{p}'$}}
\newcommand{\rp}{\mbox{$\vec{r}'$}}
\newcommand{\kp}{\mbox{$\vec{k}'$}}
\newcommand{\e}{\mbox{$\vec{e}$}}
\newcommand{\s}{\mbox{$\vec{s}$}}
\newcommand{\Li}{{\rm Li}}
\newcommand{\lsim}{\mbox{\raisebox{-0.3ex}{%
\footnotesize $\:\stackrel{<}{\sim}\:$}} }
\newcommand{\gsim}{\mbox{\raisebox{-0.3ex}{%
\footnotesize $\:\stackrel{>}{\sim}\:$}} }
\newcommand{\lb}{\left (}
\newcommand{\rb}{\right )}
\newcommand{\ep}{\epsilon}
\newcommand{\vep}{\epsilon}
\newcommand{\dd}{{\rm d}}
\newcommand{\om}{\omega}

\newcommand{\sS}{\mbox{$\vec{\sigma}\vec{\sigma}'$}}
\newcommand{\si}{\mbox{$\vec{\sigma}$}}
\newcommand{\vgamma}{\mbox{$\vec{\gamma}$}}
\newcommand{\vxi}{\mbox{$\vec{\xi}$}}

\newcommand{\pop}[1]{\mbox{$\Lambda_+(#1)$}}
\newcommand{\nep}[1]{\mbox{$\Lambda_-(#1)$}}
\newcommand{\Dafne}{DA$\Phi$NE}
\newcommand{\mar}{\marginpar{***}}
\def\spab#1.#2.#3{\langle\mskip-1mu{#1}
                  | #2 | {#3}\mskip-1mu]}
\def\spba#1.#2.#3{[\mskip-1mu{#1}
                  | #2 | {#3}\mskip-1mu\rangle}
\def\spa#1.#2{\langle#1\,#2\rangle}
\def\spb#1.#2{[#1\,#2]}

\def\dk#1#2#3{
\begin{array}{r c l}
#1 & \rightarrow & #2 \\
 & & \bentarrow #3
\end{array}
}

%% file: ALLcoefftable.tex
 \begin{math}
 \begin{array}{| l || l || l |}
 \hline
  & (1^+,2^+) & (1^-,2^+) \\
 \hline
           d_2^{d=6} &    0                                           &   -0.1629463101\times 10^{1}   -0.1341162858\times 10^{2}i \\
 \hline
           d_3^{d=6} &    0                                           &   +0.1962361656\times 10^{1}   +0.1557616902\times 10^{1}i \\
 \hline
           d_1^{(0)} &    +0.6333405849\times 10^{-1}   +0.1226578399i&   0                     \\
           d_1^{(2)} &    -0.4596242026\times 10^{-1}   +0.8702498187\times 10^{-1}i&   -0.2581442463   +0.1155542839\times 10^{1}i \\
           d_1^{(4)} &    +0.1928295436   -0.7588372864\times 10^{-1}i&   -0.1682541178   -0.1209633209i \\
 \hline
           d_2^{(0)} &    0                                           &   0                     \\
           d_2^{(2)} &    -0.4441112592\times 10^{1}   +0.9753053731\times 10^{1}i&   -0.3067119777\times 10^{2}   -0.1290256006\times 10^{2}i \\
           d_2^{(4)} &    +0.1928295436   -0.7588372864\times 10^{-1}i&   -0.1682541178   -0.1209633209i \\
 \hline
           d_3^{(0)} &    0                                &   0                     \\
           d_3^{(2)} &    +0.8583748171   -0.8394523120i&   -0.1354800014\times 10^{-1}   +0.1143682787\times 10^{1}i \\
           d_3^{(4)} &    +0.1928295436   -0.7588372864\times 10^{-1}i&   -0.1836679557   +0.9595652764\times 10^{-1}i \\
 \hline
 \hline
           c_1^{(0)} &    0                                &   0                     \\
           c_1^{(2)} &    +0.3361485887   +0.6802422581i&   -0.1121703089\times 10^{1}   +0.2624217171i \\
 \hline
           c_2^{(0)} &    0                                &   +0.7348718900   -0.1313014150i \\
           c_2^{(2)} &    -0.2538256662   +0.9988738044\times 10^{-1}i&   -0.9366786568   -0.1462703615i \\
 \hline
           c_3^{(0)} &    +0.3507990156\times 10^{-1}   +0.6793856342\times 10^{-1}i&   0                     \\
           c_3^{(2)} &    +0.1039551379   -0.2479536910i&   +0.1146266565\times 10^{1}   +0.2249571672\times 10^{-1}i \\
 \hline
           c_4^{(0)} &    +0.1121717915\times 10^{-1}   -0.2172409281\times 10^{-1}i&   0                     \\
           c_4^{(2)} &    -0.5458186613\times 10^{-1}   -0.3857256335\times 10^{-1}i&   +0.7117245915\times 10^{-1}   +0.5536768984\times 10^{-2}i \\
 \hline
           c_5^{(0)} &    +0.1353203319\times 10^{-1}   +0.2620722562\times 10^{-1}i&   0                     \\
           c_5^{(2)} &    -0.7832350223\times 10^{-1}   +0.2515373110\times 10^{-1}i&   -0.2736896597\times 10^{-1}   +0.2057525425\times 10^{-1}i \\
 \hline
           c_6^{(0)} &    +0.3739475560\times 10^{-1}   -0.7242169624\times 10^{-1}i&   0                     \\
           c_6^{(2)} &    +0.1737216593   +0.4034138715i&   +0.9275509368   -0.1094743189\times 10^{1}i \\
 \hline
 \hline
                 b_1 &    0                                                         &   +0.2169627135   +0.6541598060i \\
 \hline
                 b_2 &    +0.3181092629\times 10^{-2}   +0.2424812667\times 10^{-1}i&   -0.4358650717\times 10^{-1}   -0.1107093424i \\
 \hline
                 b_3 &    +0.7631131534\times 10^{-2}   +0.1018659988\times 10^{-1}i&   -0.7960623716\times 10^{-1}   -0.1441385947i \\
 \hline
                 b_4 &    +0.6583925572\times 10^{-2}   -0.1993931235\times 10^{-1}i&   -0.7083144889\times 10^{-1}   -0.3686048632i \\
 \hline
                 b_5 &    -0.1739614974\times 10^{-1}   -0.1449541420\times 10^{-1}i&   -0.2293852032\times 10^{-1}   -0.3070700563\times 10^{-1}i \\
 \hline
 \hline
                   R &    +0.1713240385\times 10^{-1}   +0.1341860496i&   +0.6830612977\times 10^{-1}   +0.9213040514\times 10^{-1}i \\
 \hline
 \end{array}
 \end{math}